\documentclass[aps, prd, 10pt, twocolumn, superscriptaddress,noshowpacs, preprintnumbers
,nofootinbib,floatfix]{revtex4-2}

\usepackage[dvipsnames]{xcolor}
\usepackage[colorlinks=true,breaklinks=true]{hyperref}
\hypersetup{allcolors=[rgb]{0.0 0.0 0.70},linkcolor=blue}
\usepackage{orcidlink}
\usepackage{microtype}

\usepackage{amsmath}
\usepackage{amsfonts}
\usepackage{amssymb}
\usepackage{mathtools}
\usepackage{multirow}
\usepackage{bm}
\usepackage[utf8]{inputenc}

\usepackage{graphicx}
\usepackage{dcolumn}
\usepackage{bm}
\usepackage{comment}

\usepackage{multirow}
\usepackage{booktabs}
\usepackage[normalem]{ulem}

\begin{document}

\title{Flavor Conversion Enhances or Suppresses  Supernova Explodability\\ Independent of the Progenitor Mass}

\author{Mariam Gogilashvili \orcidlink{0000-0002-6944-8052}}
 \email{mariam.gogilashvili@nbi.ku.dk}
 
\author{Irene Tamborra \orcidlink{0000-0001-7449-104X}}
 \email{tamborra@nbi.ku.dk}
\affiliation{Niels Bohr International Academy and DARK, Niels Bohr Institute, University of Copenhagen, Blegdamsvej 17, 2100, Copenhagen, Denmark
}

\begin{abstract}
Flavor conversion can affect the neutrino-driven delayed explosion mechanism of collapsing massive stars, altering the efficiency of shock revival. We perform core-collapse supernova  simulations in spherical symmetry for a set of progenitors with masses of $9.75\, M_\odot$, $11\, M_\odot$,  $16.5\, M_\odot$,  $28\, M_\odot$,  $40\, M_\odot$, and  $60\, M_\odot$, accounting for  a mixing-length treatment for convection. Flavor conversion is modeled assuming  instantaneous flavor equipartition below a critical baryon density, while conserving the lepton number. Regardless of the progenitor compactness, its mass, or the nuclear equation of state, we find that flavor conversion can increase  heating (cooling) and enhance (hinder) the supernova explosion, if triggered near the gain (neutrino decoupling) region. Our findings suggest that the interplay among the region of the supernova core  where flavor conversion occurs, the progenitor properties, and the nuclear equation of state is crucial in determining the fate of explosion and the properties of the compact remnant.
\end{abstract}

\maketitle

\section{\label{sec:Intro}Introduction}
The end of the life of  massive stars is  among the 
most extreme phenomena in our Universe and can lead   to energetic explosions known as core-collapse supernovae (CCSNe). The latter play a central role driving turbulence in the interstellar medium and synthesizing many of the chemical elements in our Universe~\cite{woosley2002, DIEHL2021,low2004,Janka2025,Burrows2021}.  
When the iron core of a massive star becomes gravitationally unstable, it collapses on a dynamical timescale. As the  core reaches nuclear densities, the stiffening of the nuclear equation of state halts the collapse, launching a shock wave that propagates outward through the infalling stellar material. However, the shock rapidly loses energy, through  dissociation of heavy nuclei and neutrino losses, and 
stalls~\cite{HILLEBRANDT1981, MAZUREK82, MAZUREK1982}. 
The subsequent evolution of the stalled shock determines the fate of the star. If the shock is revived and expands outward,  a successful explosion occurs, with the remnant compact object being a neutron star~\cite{BURROWS1986}. If the shock fails to be revived, continued accretion onto the proto-neutron star can drive the latter beyond its maximum mass, leading to collapse into a black hole~\cite{FISCHER2009, OCONNOR2011,Burrows2025}. 

Shock revival in most collapsing massive stars is expected to occur via the delayed neutrino-heating mechanism~\cite{COLGATE1966, BETHE1985}. In this scenario, neutrinos emitted from the hot proto–neutron star deposit energy in the region behind the stalled shock, known as the gain region. If the heating rate is sufficiently strong, the pressure in the gain layer can increase enough to push the shock outward and trigger its further expansion. Numerical simulations of CCSNe suggest that, in addition to neutrino heating, hydrodynamical instabilities (e.g.,  neutrino-driven convection and the standing accretion shock instability) can enhance the efficiency of neutrino heating and aid the explosion~\cite{Bethe1990RvMP,Blondin2003,Blondin2007,Foglizzo2012,Fernandez2010,Janka2025,Janka2016,Burrows2007}.  
Neutrinos therefore play a central role in the explosion mechanism, and small changes in neutrino spectral properties  can significantly alter the heating conditions in the gain region, affecting the fate of the CCSN explosion ~\cite{JANKA2012,Raffelt2025,Tamborra2025}.

State-of-the-art simulations of CCSNe include sophisticated treatments of neutrino transport and neutrino–matter interactions (see, e.g., Refs.~\cite{Mezzacappa2020,Fischer2024} for an overview). However, they generally neglect neutrino flavor conversion, except for a few preliminary attempts~\cite{Strack2005, Dasgupta2012}. This omission was justified by the understanding that flavor conversion could occur only beyond the gain layer, therefore having little impact on the explosion mechanism~\cite{Dasgupta2012}.

Recent advances in neutrino self-interaction physics suggest that flavor conversion may already take place during neutrino decoupling from matter~\cite{Sawyer2005,Sawyer2016,Chakraborty2016,Izaguirre2017,Shalgar2023,Johns2023,Fiorillo2025}. In the dense environment of the CCSN core, neutrino–neutrino refraction can dominate the flavor evolution (see, e.g., Refs.~\cite{Duan2010,Mirizzi2016,Tamborra2021,Volpe2024,Johns2025}). Since flavor conversion can significantly modify the spectral and flavor composition of the neutrino field, it could in turn alter the charged-current absorption rates that drive neutrino heating behind the stalled shock, with important implications for shock revival.

Despite its potential importance, the integration of  neutrino flavor conversion into hydrodynamical CCSN simulations presents major theoretical and computational challenges. The characteristic scales entering the neutrino kinetic equations can be vastly different; moreover, the neutrino flavor field is expected to evolve on temporal and spatial scales orders of magnitude smaller than the resolution commonly employed  in global CCSN simulations. Although rapid progress has been made in solving the neutrino kinetic equations within simplified frameworks~\cite{Nagakura2022,Shalgar2023,Shalgar2023a,Xiong2024,Cornelius2024,Shalgar2025}, the full seven-dimensional solution and its coupling to the CCSN hydrodynamics remain beyond the current computational capabilities.

To bridge the gap between the microscopic physics of neutrino flavor evolution and the macroscopic CCSN dynamics, subgrid models have been proposed to predict the flavor outcome without solving the kinetic equations~\cite{Xiong2021,Just2022,Padilla2022,Ehring2023,Zaizen2023,Zaizen2023a,Nagakura2024,Goimil2025,Liu2025,Xiong2025,Johns2025_subgrid}. These schemes typically employ  symmetry assumptions to simplify the problem, but offer the key advantage of being potentially easy to implement  into hydrodynamical simulations.
Recent work relies on these subgrid models to assess the impact of flavor conversion on the explosion mechanism, accounting for an estimated quasi-steady-state flavor configuration in hydrodynamical simulations~\cite{Ehring2023, Ehring2023abs,Nagakura2023,Wang2025,Akaho2026,Mori2025}. The results indicate a mass-dependent effect: flavor conversion can enhance explosions in low-mass progenitors ($\lesssim 12 \,M_\odot$) while potentially suppressing explosions in high-mass models ($\gtrsim 20 \, M_\odot$)~\cite{Ehring2023abs, Nagakura2023,Akaho2026,Mori2025}. This trend has been attributed to the fact that flavor conversion favors shock expansion when the accretion rate is low, but may hinder the explosion under conditions of high accretion. 

In this paper, we explore the consequences of neutrino flavor conversion on the CCSN explodability. We 
incorporate the flavor conversion scheme of Ref.~\cite{Ehring2023} into the open-source, general relativistic radiation hydrodynamics simulation \texttt{GR1D}~\cite{OConnor2010, OConnor2015}. We perform spherically symmetric simulations of pre-CCSN progenitors with masses of $9.75\, M_\odot$, $11\, M_\odot$,  $16.5\, M_\odot$,  $28\, M_\odot$,  $40\, M_\odot$, and  $60\, M_\odot$, using the Steiner, Fischer, and Hempel equation of state (SFHo EOS)~\cite{STEINER2013} and the Lattimer and Swesty  nuclear equation of state with incompressibility modulus equal to $220$~MeV (LS220 EOS)~\cite{LATTIMER1991}. Hydrodynamics and neutrino transport are evolved in spherical symmetry, whereas the effects of neutrino-driven convection are taken into account through a mixing-length treatment~\cite{Boccioli2021_STIR_GR}. We find that the empirical bifurcation of the CCSN explodability, depending on the progenitor mass and reported in Refs.~\cite{Ehring2023abs,Akaho2026}, is the result of a subtle interplay between the  microphysics (such as neutrino flavor conversion and EOS) and  progenitor structure (i.e., the core compactness). However,  depending on the spacial region in the CCSN interior where flavor conversion takes place,  our results show that the latter can change the CCSN fate  regardless of the progenitor mass. 

The structure of this paper is as follows. In Sec.~{\ref{sec:Methods}}, we describe the numerical setup for our simulation and
 the schematic approach adopted to model flavor conversion. We present our results in Sec.~\ref{sec:Results}:  we explore the impact  of the radial range where flavor conversion is triggered on the CCSN explodability; additionally, we investigate the role of the EOS and the effects of flavor conversion for progenitors with different masses. We summarize and discuss our findings in Sec.~\ref{sec:D&C}. 

\section{\label{sec:Methods} Problem setup}
In this section, we introduce the \texttt{GR1D} simulation framework used to model the stellar collapse. We also outline the method used to model flavor conversion in the CCSN core. 
\subsection{\label{subsec:GR1D} Supernova simulations}

To follow the collapse of the iron core and its post-bounce evolution, we employ the open-source, spherically symmetric code \texttt{GR1D}~\cite{OConnor2010, OConnor2015}. \texttt{GR1D} solves the equations of general relativistic hydrodynamics in spherical symmetry and is coupled to an energy-dependent neutrino transport module. 

The hydrodynamical evolution is formulated in conservative form for a relativistic fluid. 
The spatial discretization of the hydrodynamical equations employs a finite-volume scheme, and the time integration is performed using an explicit second-order Runge--Kutta method.
We perform the hydrodynamic evolution on a spherically symmetric radial grid consisting of $600$ zones. The innermost region is resolved with a uniform  spacing of $3 \times 10^4$~cm  up to $20$~km, while the grid spacing increases logarithmically at larger radii.

In CCSNe, multi-dimensional effects--such as neutrino-driven convection and turbulence--play a crucial role in aiding successful explosions as they increase the dwell time of accreting matter in the gain region,  enhancing the efficiency of neutrino heating. To account for these effects, we perform 1D+ simulations with \texttt{GR1D}, using the Supernova Turbulence in Reduced-dimensionality (\texttt{STIR}) model ~\cite{Boccioli2021_STIR_GR}. In this approach, neutrino-driven convection and turbulence are captured through a Reynolds decomposition of the hydrodynamic variables; the resulting averaged equations are evolved using a time-dependent mixing-length theory~\cite{Couch2020_STIR}.

The stability of the fluid to convection is determined by the Brunt--V\"ais\"al\"a frequency, which depends on the local thermodynamic gradients and gravitational acceleration. In regions where this quantity indicates instability, turbulent motions are expected to develop.  To account for such turbulent motions, \texttt{STIR} estimates their characteristic strength using a mixing-length prescription. The latter is characterized by the dimensionless parameter  $\alpha_{\rm MLT}$, which scales the mixing length to the pressure scale height  
($\Lambda = \alpha_{\rm MLT} H_P$, 
with $H_P = P/\rho g$ being the pressure scale height, $P$ being the pressure, $\rho$ the density, and $g$ the gravitational acceleration). 
The parameter $\alpha_{\rm MLT} = \mathcal{O}(1)$ sets the efficiency of convective transport. 
In the following, we use  $\alpha_{\rm MLT}=1.51$, unless otherwise specified.

Neutrino transport is treated using an energy-dependent moment formalism~\cite{SHIBATA2011, CARDALL2013}. 
The hierarchy of moment equations is closed with an analytic M1 closure relation to approximate the radiation pressure tensor in terms of lower-order moments.   
The neutrino spectrum is discretized using $18$ energy bins and includes three species: 
$\nu_e$, $\bar{\nu}_e$, and $\nu_x = \bar\nu_x$ (with $x=\mu$ or $\tau$).
Neutrino--matter interaction (emission, absorption, and scattering processes) rates are modeled using \texttt{NuLib}~\cite{OConnor2015}.

We adopt progenitor models from Ref.~\cite{Sukhbold2016}. We select a set of  stellar models with zero-age main sequence masses of $9.75\, M_\odot$, $11\, M_\odot$,  $16.5\, M_\odot$,  $28\, M_\odot$,  $40\, M_\odot$, and  $60\, M_\odot$, that span a range of core properties and compactness. 
To assess the sensitivity of our results to the CCSN microphysics, we perform simulations using two EOSs: SFHo~\cite{STEINER2013} and LS220~\cite{LATTIMER1991}. 

\subsection{\label{subsec:FC} Flavor conversion scheme}
Flavor conversion in the CCSN core depends on the shape of the angular distribution of the electron-neutrino lepton number.   
Due to their high computational cost, most state-of-the-art neutrino hydrodynamical simulations of CCSNe evolve only the lowest angular moments of the (anti)neutrino distribution functions, with a few notable exceptions~\cite{Fischer2010, Sumiyoshi2012, Tamborra2017, Nagakura2018, Akaho2021, Akaho2026}. As a result, one needs to reconstruct the neutrino angular distributions in post-processing. This reconstruction is inherently limited, since only the lowest-order moments are available~\cite{Johns2021, Cornelius2025a, Akaho2026}. Consequently, subgrid prescriptions for neutrino flavor conversion (e.g., those proposed in Refs.~\cite{Xiong2021, Padilla2022, Zaizen2023, Zaizen2023a, Nagakura2024, Goimil2025, Liu2025, Xiong2025, Johns2025_subgrid}) cannot yet provide reliable predictions when applied to moment-based CCSN simulations. This limitation also applies to our set of CCSN models.

To assess the impact of flavor conversion on the explosion outcome, we adopt the parametric scheme of Refs.~\cite{Just2022,Ehring2023}. We assume that a quasi-steady flavor configuration is established on timescales shorter than the CCSN dynamical ones. We then evolve the neutrino and antineutrino flavors toward equipartition, while conserving the lepton number. This approximation captures the largest impact of flavor conversion that one could expect.

We follow the same procedure described in Ref.~\cite{Ehring2023}. For each $i$--$\mathrm{th}$ energy bin, we enforce conservation of the electron lepton number, $\mathcal{L}_i$, before and after flavor equipartition is achieved; hence 
$\mathcal{L}_{i} = n_{\nu_e,i} - n_{\bar{\nu}_e,i}
= n'_{\nu_e, i} - n'_{\bar{\nu}_e, i}
= \mathcal{L}'_{i}$.  This approach neglects the fact that neutrino self-interactions  
may take place between  neutrinos and antineutrinos with different momenta, redistributing the total ELN across momentum bins. Tracking this cross-momentum coupling, however, would require a statistical treatment and the computation of the  neutrino angular distributions, which is not possible within the moment-based transport scheme of \texttt{GR1D}. Therefore, enforcing group-wise ELN conservation provides the simplest and computationally tractable subgrid closure that allows to assess the qualitative impact of  flavor conversion on neutrino heating while preserving the initial spectral hierarchy. 

We also conserve the total number of neutrinos and antineutrinos. This gives $n'_{\nu_e, i} + 2\ n'_{\nu_x, i} = n_{\nu_e, i} + 2\ n_{\nu_x, i}$ and $n'_{\bar{\nu}_e, i} + 2\ n'_{{\nu}_x, i} = n_{\bar{\nu}_e, i} + 2\ n_{{\nu}_x, i}$.
Under these constraints, flavor equipartition yields the following number densities:
\begin{eqnarray}
\label{eq:equip}
n'_{\nu_e, i} &=& n_{\mathrm{eq}, i} + \mathrm{max}(0, \mathcal{L}_{i})\, ,\\
\label{eq:equip1}
n'_{\bar\nu_e, i} &=& n_{\mathrm{eq}, i} + \mathrm{max}(0, - \mathcal{L}_{i})\, ,\\
\label{eq:equip2}
n'_{\nu_x, i} &=& n_{\mathrm{eq}, i} \, .
\end{eqnarray}
Here, $n_{\mathrm{eq}, i}$ follows from number conservation and depends on the sign of $\mathcal{L}_{i}$: 
\begin{eqnarray}
n_{\mathrm{eq}, i} = \begin{cases}
n'_{\bar{\nu}_e, i} = n'_{{\nu}_x, i} = 
\dfrac{1}{3}\left(n_{\bar{\nu}_e,i} + 2 n_{\nu_x,i}\right)  \text{if } \mathcal{L}_{i} > 0\, , \\
n'_{{\nu}_e,i} =  n'_{{\nu}_x,i} =
\dfrac{1}{3}\left(n_{\nu_e,i} + 2 n_{\nu_x,i}\right)  \text{if } \mathcal{L}_{i} \le 0\, .
\end{cases}
\end{eqnarray}

We also conserve the total momentum of neutrinos and antineutrinos in each energy bin. This gives $\mathbf{F}_{\nu_e, i}+ \mathbf{F}_{\bar\nu_e, i}+ 4\ \mathbf{F}_{\nu_x, i} = \mathbf{F}'_{\nu_e, i}+\mathbf{F}'_{\bar\nu_e, i}+ 4\ \mathbf{F}'_{\nu_x, i}$. In the  regions where $\nu_e,\ \bar\nu_e \rightarrow \nu_x$,   the fluxes read
\begin{eqnarray}
\label{eq:flux}
\mathbf{F}'_{\nu_e, i} &=& \mathbf{F}_{\nu_e, i} \frac{n_{\nu_e, \mathrm{eq}, i}}{n_{\nu_e, i}}\, , \\
\mathbf{F}'_{\bar\nu_e, i} &=& \mathbf{F}_{\bar\nu_e, i} \frac{n'_{\bar\nu_e, i}}{n_{\bar\nu_e, i}}\, , \\
\label{eq:flux1}
\mathbf{F}'_{\nu_x, i} &=& \mathbf{F}_{\nu_x, i} +\frac{\mathbf{F}_{\nu_e, i} [1-(n'_{\nu_e, i}/n_{\nu_e, i})]}{4}\\\nonumber &+& \frac{\mathbf{F}_{\bar\nu_e, i} [1-(n'_{\bar\nu_e, i}/n_{\bar\nu_e, i})]}{4}\, . 
\label{eq:flux2}
\end{eqnarray}
On the other hand,  where $\nu_x \rightarrow \nu_e,\ \bar\nu_e$,  the fluxes become
\begin{eqnarray}
\label{eq:flux3}
\mathbf{F}'_{\nu_e, i} &=& \mathbf{F}_{\nu_e, i}+2 [1 - ({n'_{\nu_x, i}}/{n_{\nu_x, i}})] \mathbf{F}_{\nu_x, i} \, , \\
\label{eq:flux4}
\mathbf{F}'_{\bar\nu_e, i} &=& \mathbf{F}_{\bar\nu_e, i}+2 [1 - ({n'_{\nu_x, i}}/{n_{\nu_x, i}})] \mathbf{F}_{\nu_x, i} \, , \\
\label{eq:flux5}
\mathbf{F}'_{\nu_x, i} &=& \mathbf{F}_{\nu_x, i} \frac{n'_{\nu_x, i}}{n_{\nu_x, i}}\, .
\end{eqnarray}
This scheme respects Pauli blocking by ensuring that the occupation number in each energy bin never exceeds the fermionic phase-space limit. For further details, we refer the reader to Sec.~II.C of Ref.~\cite{Ehring2023}.

Global simulations of neutrino kinetics show that flavor equipartition can occur in the neutrino or antineutrino sector. However, it is not a generic outcome~\cite{Shalgar2024, Shalgar2025}, it depends on the angular configurations of the (anti)neutrino distributions. In case of slow flavor conversion, which is also expected in the CCSN core, equipartition may not be reached~\cite{Shalgar2025, Fiorillo2025, Padilla2025}.

From a moment-based neutrino scheme, we cannot reliably identify the regions nor the exact post-bounce time intervals where flavor conversion takes place~\cite{Johns2021, Cornelius2025a}. 
Therefore, similarly to Ref.~\cite{Ehring2023}, we apply Eqs.~(\ref{eq:equip})--(\ref{eq:equip2}) and (\ref{eq:flux})--(\ref{eq:flux5}) for each energy bin, at every time step shortly after bounce ($0.02\, \mathrm{s}$ after bounce~\footnote{For numerical stability, the flavor-conversion scheme is activated at $t = 0.02\, \mathrm{s}$ after core bounce rather than at bounce itself. However, we have verified that our findings are not sensitive to   changes of the activation time near the bounce.}).
We also define a threshold matter density, $\rho_c$, which we vary arbitrarily from $10^{13}$ to $10^9\, \mathrm{g/cm^3}$, covering the radial region between the neutrinosphere and the stalled shock. Flavor conversion is activated in zones where $\rho < \rho_c$. In these regions, the (anti)neutrino distributions are instantaneously modified according to Eqs.~(\ref{eq:equip}), (\ref{eq:equip1}), and (\ref{eq:flux})--(\ref{eq:flux5})\footnote{This scheme is applied as an
operator-split correction after the implicit neutrino-matter source
update. Hence it does not exactly capture the stiff, simultaneous coupling
between matter and flavor equilibration expected at high $\rho_c$. We have carried out  a
timestep-sensitivity test for 
$\rho_c=10^{13}\,\mathrm{g/cm^3}$; we find that  the operator splitting error has a
negligible ($\lesssim 1$--$3~\%$) effect on the shock radius evolution
and does not alter our conclusions.}.

\section{Results: Impact of flavor conversion}\label{sec:Results}

In this section, we present our main findings. We first focus on the
progenitor with mass of $16.5\,M_\odot$. We explore how the shock evolution and explosion fate 
depend on the location where flavor conversion takes place. 
Then, we examine how the effects of flavor conversion depend on the choice of the EOS.
Finally, we analyze our set of CCSN models to understand whether the impact of flavor conversion on the CCSN explodability depends on differences in the progenitor structure, mass, and compactness.

\subsection{Impact on the explosion fate}\label{sec:diff_rhoc}

\begin{figure}
    \centering
    \includegraphics[width=0.48\textwidth]{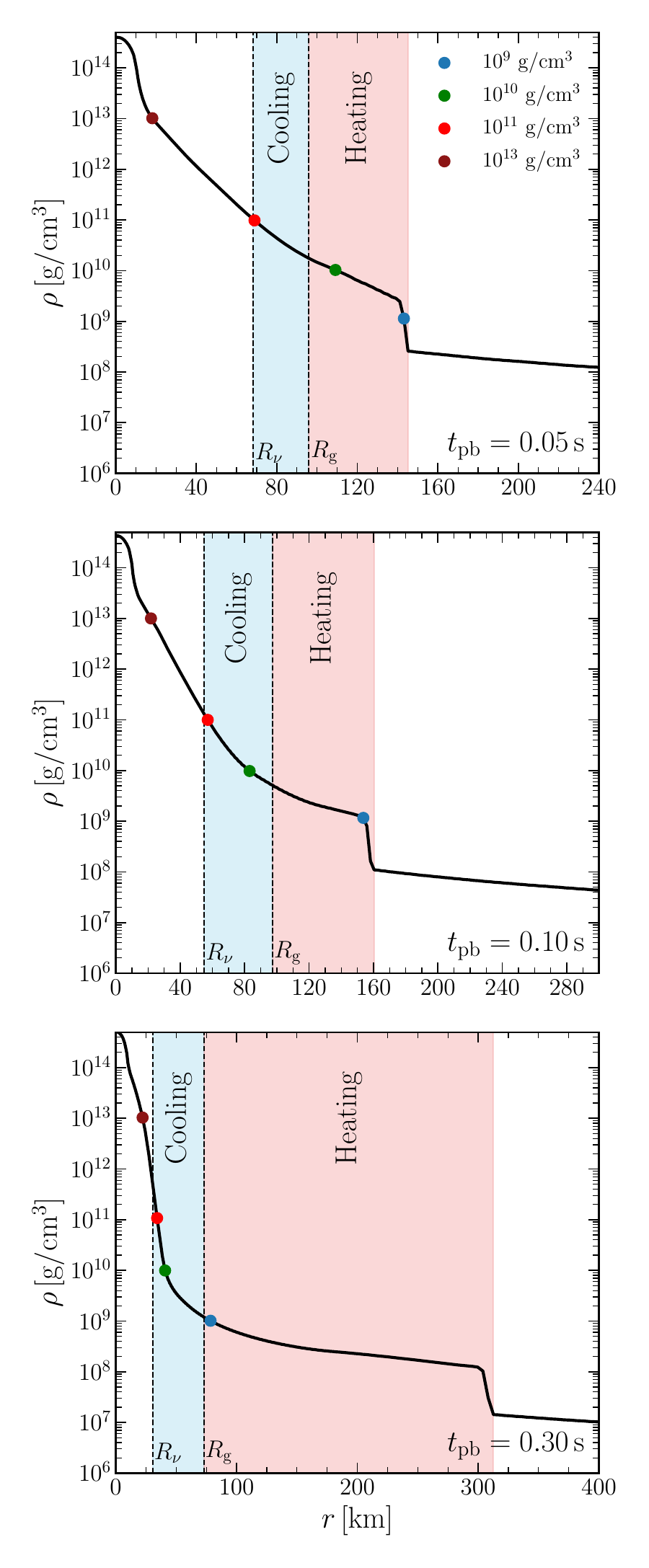}
    \caption{Radial profiles of the baryon density  for our reference CCSN model with mass of $16.5\, M_\odot$ and SFHo EOS, extracted at different post-bounce times: $t_{\mathrm{pb}} = 0.05 \, , 0.1 \, , \mathrm{and} \, 0.3\, \mathrm{s}$, from top to bottom respectively. The shaded regions mark the cooling (blue) and gain (red) regions behind the shock front. The dashed vertical lines indicate the neutrinosphere ($R_\nu$) and gain radii ($R_\mathrm{g}$). Colored dots correspond to selected values of the critical density, $\rho_c$, below which flavor conversion is triggered. The baryon density profile evolves as a function of post-bounce time, and therefore the regions where flavor conversion occurs change accordingly.
}
    \label{fig:density_snaps}
\end{figure}

\begin{figure*}
    \centering    \includegraphics[width=0.9\textwidth]{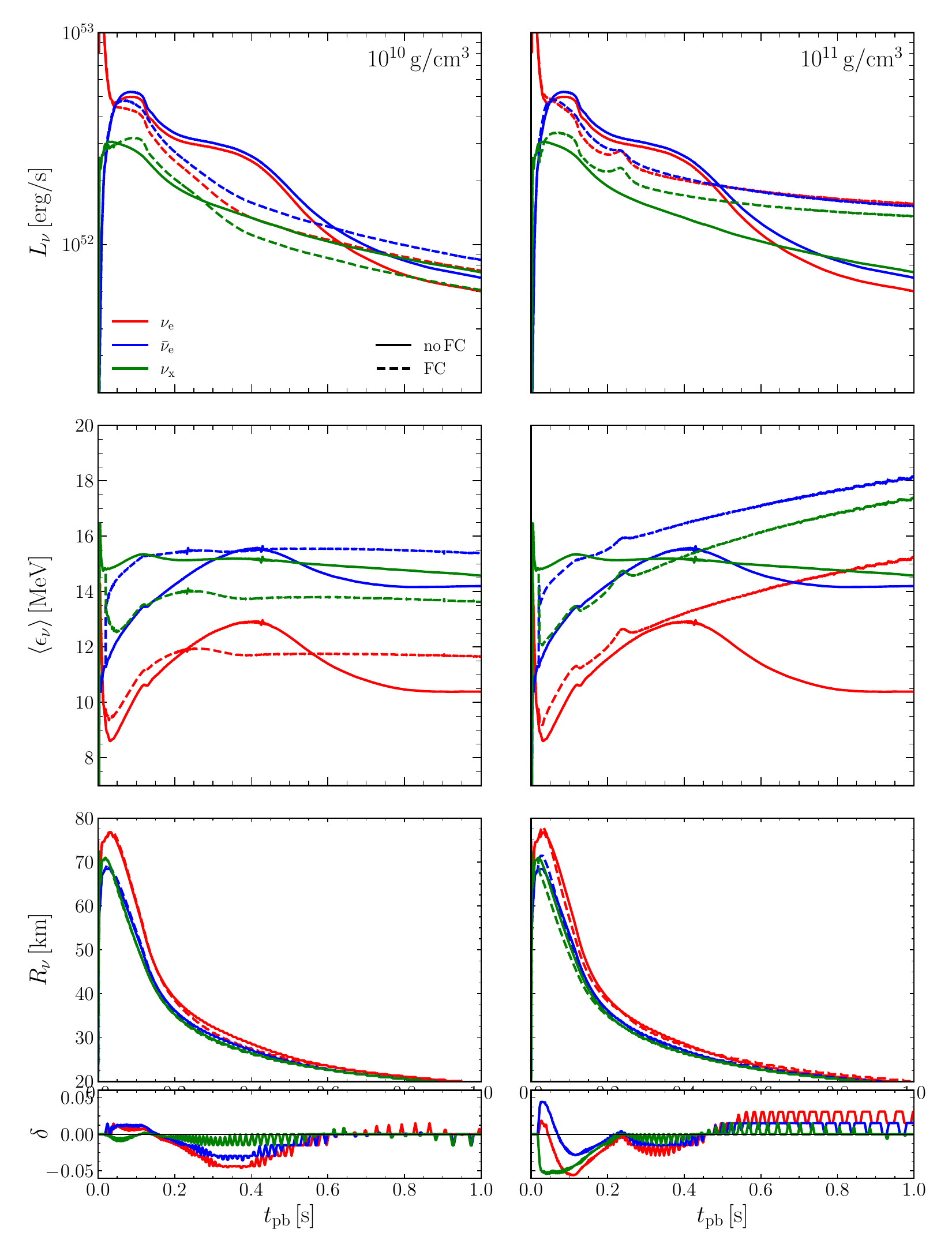}
    \caption{Evolution of the neutrino luminosity, average energy (both extracted at 
    $500\, \mathrm{km}$ and measured in the lab frame) and energy-averaged neutrinosphere radius as functions of the post-bounce time, from top to bottom respectively, for our benchmark CCSN model with mass of $16.5\, M_\odot$ and SFHo EOS. 
    The different colors denote the neutrino species: $\nu_e$ (red), $\bar{\nu}_e$ (blue), and $\nu_x$ (green). The solid lines correspond to the CCSN model without flavor conversion (no FC), whereas the dashed lines show the cases where flavor equipartition is imposed below a critical density (FC, $\rho_c = 10^{10}\, \mathrm{g/cm{^3}}$ on the left and  $\rho_c = 10^{11} \, \mathrm{g/cm^3}$ on the right). In order to highlight the variations of the neutrinosphere radius, the bottom panels display the relative variation 
    $\delta={(R_{\nu_{\mathrm{FC}}} - R_{\nu_{\mathrm{no\, FC}}})}/{R_{\nu_{\mathrm{no\, FC}}}}$ (note that the small spikes observable in  $\delta$ are a numerical artifact due to the time binning).  Flavor conversion affects the neutrino emission properties, enhancing (suppressing)  heating for $\rho_c = 10^{10} \, \mathrm{g/cm^3}$ ($\rho_c = 10^{11}\, \mathrm{g/cm^3}$).}
    \label{fig:nuprops}
\end{figure*}

To investigate the effects of flavor conversion on CCSN explosions, we focus on the $16.5\, M_\odot$ progenitor simulated using the SFHo EOS. We vary the critical density, $\rho_c$, from $10^{9}$ to $10^{13}\, \mathrm{g/cm^3}$, spanning regions from the proto-neutron star interior to the gain layer behind the shock. 
Figure~\ref{fig:density_snaps} shows the baryon density profiles for the $16.5\, M_\odot$ progenitor without flavor conversion at three different post-bounce times: $t_{\rm pb} = 0.05, \, 0.1, \, \mathrm{and} \, 0.3\, \mathrm{s}$. The shaded areas indicate the cooling region (in blue) and the heating region (in red) behind the shock front. The vertical dashed lines represent the neutrinosphere radius (defined as the flavor- and energy-averaged  radius where the optical depth is $\tau = 2/3$) and the gain radius (i.e., the location where neutrino heating balances cooling). The colored dots mark the values of $\rho_c$ below which flavor conversions are triggered. As the post-bounce time increases, the density structure changes, and the radial locations corresponding to different $\rho_c$ values shift accordingly. Lower $\rho_c$ values fall near or within the gain region, intermediate values of $\rho_c$ lie in the cooling layer, and the large values of $\rho_c$ are located in the  proto-neutron star interior.
For illustrative purposes, we focus on two critical densities: $\rho_c = 10^{11} \, \mathrm{g/cm^3}$, which lies near the neutrinosphere, and $\rho_c = 10^{10} \,  \mathrm{g/cm^3}$, corresponding to radii close to the gain region and the shock.
\begin{figure*}
    \centering   
    \includegraphics[width=\textwidth]{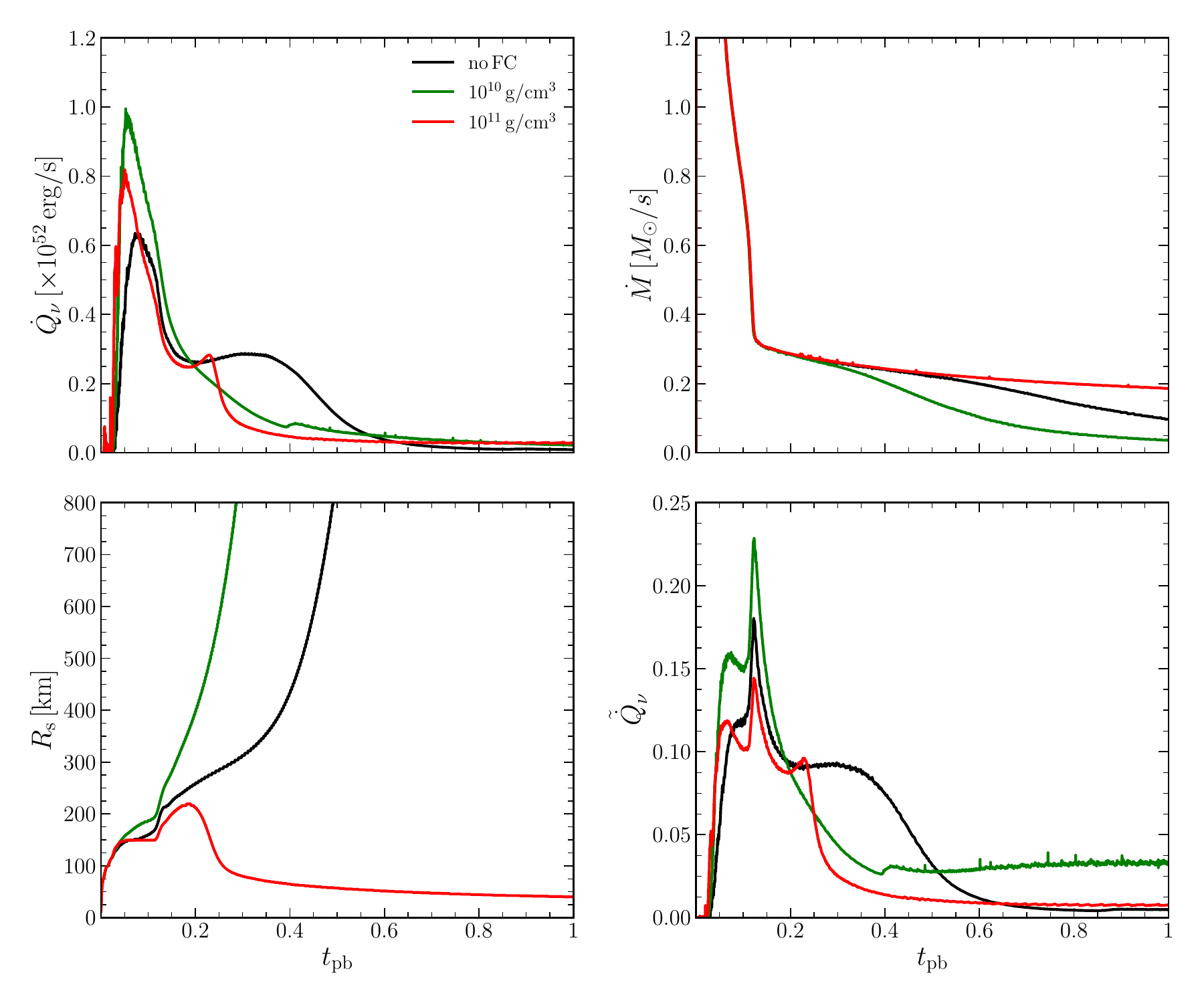} 
    \caption{Evolution of the net neutrino heating rate (top left panel), mass accretion rate (top right panel), shock radius (bottom left panel) and the normalized net heating rate in the gain region (Eq.~\ref{eq:tildeQ}, bottom right panel) as functions of the post-bounce time for the $16.5\, M_\odot$ progenitor with SFHo EOS. The black line shows the reference simulation without flavor conversion (no FC), while the green and red curves correspond to flavor conversion being triggered at  $\rho_c = 10^{10}\,\mathrm{g/cm^3}$ and $\rho_c = 10^{11}\,\mathrm{g/cm^3}$, respectively. The CCSN model with $\rho_c = 10^{10}\,\mathrm{g/cm^3}$ exhibits enhanced neutrino heating and shock expansion compared to the reference CCSN model without flavor conversion, whereas the $\rho_c = 10^{11}\,\mathrm{g/cm^3}$ model shows initial enhancement of heating, but an overall weaker heating rate and shock stagnation at later times. The mass accretion rate is initially similar for all models, but later diverge with the $\rho_c = 10^{11}\,\mathrm{g/cm^3}$ ($\rho_c = 10^{10}\,\mathrm{g/cm^3}$) model maintaining the highest (lowest) accretion rate.}
    \label{fig:4panel}
\end{figure*}

Figure~\ref{fig:nuprops} displays the time evolution of the neutrino properties for all three flavors for our $16.5\, M_\odot$ model. From top to bottom, the panels display the luminosity, average energy, and the energy-averaged neutrinosphere radius. The left panels correspond to cases where flavor conversion is triggered at $\rho_c = 10^{10} \, \mathrm{g/cm^3}$, while the right panels show the case with $\rho_c = 10^{11}\, \mathrm{g/cm^3}$. 

The critical density $\rho_c = 10^{10}\, \mathrm{g/cm^3}$ (left panels of Fig.~\ref{fig:nuprops}) roughly spans  the outer cooling layer through the gain region near the shock (cf.~Fig.~\ref{fig:density_snaps}).
In this case, flavor conversion reduces the average energy of heavy-lepton neutrinos. This energy is transferred to electron neutrinos and antineutrinos, which adjust their emission properties to enhance neutrino heating in the gain region (see the upper left panel of Fig.~\ref{fig:4panel}, green vs.~black curves). The stronger heating causes the shock to move outward  earlier, into lower-density material. As a result, the mass accretion rate decreases (upper right panel of Fig.~\ref{fig:4panel}), further supporting faster shock expansion, as shown in the bottom left panel of Fig.~\ref{fig:4panel}. Our $16.5\, M_\odot$ model explodes successfully without flavor conversion (black curve, bottom left of Fig.~\ref{fig:4panel}), but flavor conversion at  $\rho_c = 10^{10}\, \mathrm{g/cm^3}$ aids the explosion.

In contrast,  $\rho_c = 10^{11}\, \mathrm{g/cm^3}$ lies at the inner edge of the cooling layer (cf.~Fig.~\ref{fig:density_snaps}). Hence, flavor conversion occurs deeper inside the proto-neutron star, close to the neutrinosphere or inner cooling layer. In this region, the direction of flavor conversion
($\nu_x\to\nu_e,\bar\nu_e$ or $\nu_e,\bar\nu_e\to\nu_x$) varies with
neutrino energy: flavor conversion tends to proceed as $\nu_e,\bar\nu_e\to\nu_x$
at low-to-intermediate energies and as $\nu_x\to\nu_e,\bar\nu_e$ in the
high-energy tail of the spectra, cf.~also Ref.~\cite{Ehring2023}. Depleting the  low-energy part of the
$\nu_e/\bar\nu_e$ spectra suppresses their luminosity, while the surviving
or enhanced high-energy tail raises their average energy (right panels of
Fig.~\ref{fig:nuprops}).
Overall the decrease of neutrino luminosity is responsible for the reduction of net heating in the gain-region ($\propto L_\nu\langle\epsilon_\nu^2\rangle$).
As shown in the top left panel of Fig.~\ref{fig:4panel}, the heating rate initially rises slightly, but then becomes substantially weaker at later times (red vs.~black curves). Meanwhile, the mass accretion rate remains high, maintaining a stronger ram pressure on the shock (top right panel of Fig.~\ref{fig:4panel}). The combination of weaker heating and sustained accretion prevents shock revival, as shown in the bottom left panel of Fig.~\ref{fig:4panel}. Consequently, flavor conversion at this $\rho_c$ hinders the explosion and leads to shock recession.

\begin{figure}
    \centering
    \includegraphics[width=0.48\textwidth]{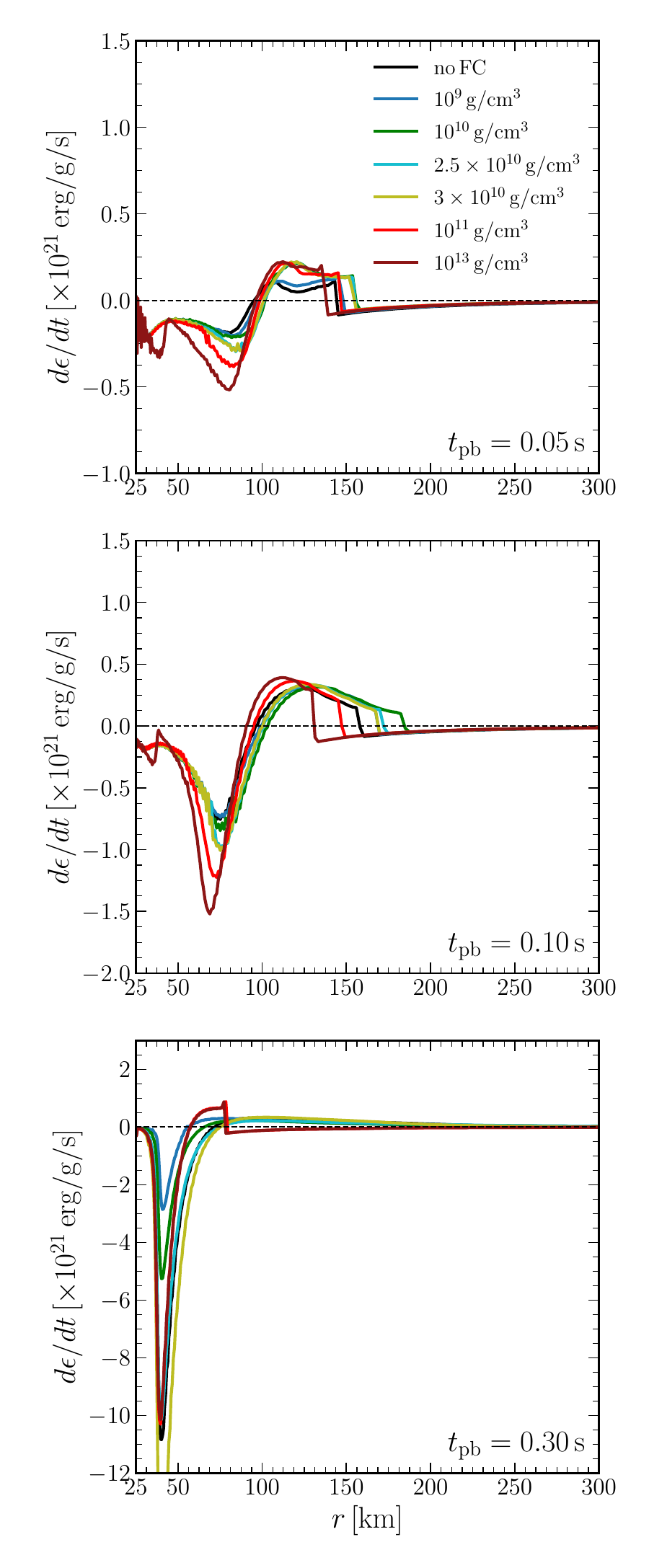}
    \caption{Radial profiles of the net specific energy change rate ($d\epsilon/dt$, neutrino heating minus cooling per unit mass) at post-bounce times of $0.05 \, \mathrm{s}$ (top), $0.1 \, \mathrm{s}$ (middle), and $0.3 \, \mathrm{s}$ (bottom). Positive (negative) values of $d\epsilon/dt$ indicate net heating (cooling). The curves of different colors correspond to different  $\rho_c$ where flavor conversion is triggered. Models with higher $\rho_c$ display enhanced cooling near the neutrinosphere and reduced  heating efficiency behind the shock. Models with  lower $\rho_c$ show increased heating efficiency in the gain region. These differences are responsible for opposite consequences on the CCSN explodability. }
    \label{fig:depsdt}
\end{figure}

More generally, whether flavor conversion favors enhanced heating or cooling depends on both the type  of conversion and the region where it occurs. Conversions from heavy-lepton to electron flavors ($x \to e$) favor  heating in the gain region, while conversions from electron to heavy-lepton neutrinos ($e \to x$) tend to reduce heating and may favor cooling. Such changes in heating affect the shock radius. Since the pre-shock
density and velocity profiles are set by the progenitor, the mass accretion
rate through the shock is thereby affected through the shock radius, $\dot{M}_s = 4\pi\rho(r_s)|v(r_s)|r_s^2$ (Fig.~3,
upper right panel): stronger heating pushes the shock outward into
lower-density infalling material, where $\dot{M}_s$ is lower, further
supporting shock expansion, while weaker heating keeps the shock at smaller
radii where $\dot{M}_s$ remains higher.
To better illustrate this trend, Fig.~\ref{fig:depsdt} shows the radial profiles of the net specific energy change rate, $d\epsilon/dt$, at three post-bounce times ($t_{\rm pb} = 0.05 \, , 0.1, \, \mathrm{and} \, 0.3\, \mathrm{s}$). This quantity represents the net neutrino heating (positive) or cooling (negative) per unit mass. The curves of different colors correspond to CCSN models with flavor conversion triggered at different  $\rho_c$, ranging from $10^{9}$ to $10^{13}\, \mathrm{g/cm^3}$.
For higher $\rho_c$ (red and brown curves), flavor conversion occurs deeper inside the proto-neutron star near the neutrinosphere, where cooling dominates and is enhanced relative to the case without flavor conversion (black curve). This increased cooling makes  the net heating behind the shock less efficient, contributing to weaker shock expansion and delayed or failed explosion.
Conversely, for lower $\rho_c$ (blue and green curves), flavor conversion occurs closer to or within the gain region, enhancing the net heating. This increased heating supports the shock expansion and favors the CCSN explosion.

A useful diagnostic quantity of the explosion conditions is the normalized net heating deposited in the gain region  introduced in Refs.~\cite{Gogilashvili2022, Gogilashvili2023, Gogilashvili2024, Boccioli2025} as a measure of the balance between neutrino heating and the competing effects of accretion and cooling:
\begin{equation}
    \tilde{\dot{Q}}_\nu = \frac{\dot{Q}_\nu R_{\mathrm{NS}}}{G\dot{M} M_{\mathrm{NS}}} \, ,
    \label{eq:tildeQ}
\end{equation}
where $G$ is the gravitational constant, $R_\mathrm{NS}$ and $M_\mathrm{NS}$ are the radius and mass of the neutron star. For our calculations, we define $R_\mathrm{NS}$ as the flavor- and energy-averaged neutrinosphere radius.
The evolution of $\tilde{\dot{Q}}_\nu$ is displayed in the bottom right panel of  Fig.~\ref{fig:4panel}. Consistent with the shock evolution, we find that the model with $\rho_c = 10^{10} \, \mathrm{g/cm^3}$ exhibits a larger average $\tilde{\dot{Q}}_\nu$ prior to the explosion compared to our CCSN model without flavor conversion, indicating more favorable conditions for the explosion. Conversely, for  $\rho_c = 10^{11}\,\mathrm{g/cm^3}$, we find an overall  smaller $\tilde{\dot{Q}}_\nu$, responsible for halting the CCSN explosion.

\begin{figure}
    \centering
    \includegraphics[width=0.48\textwidth]{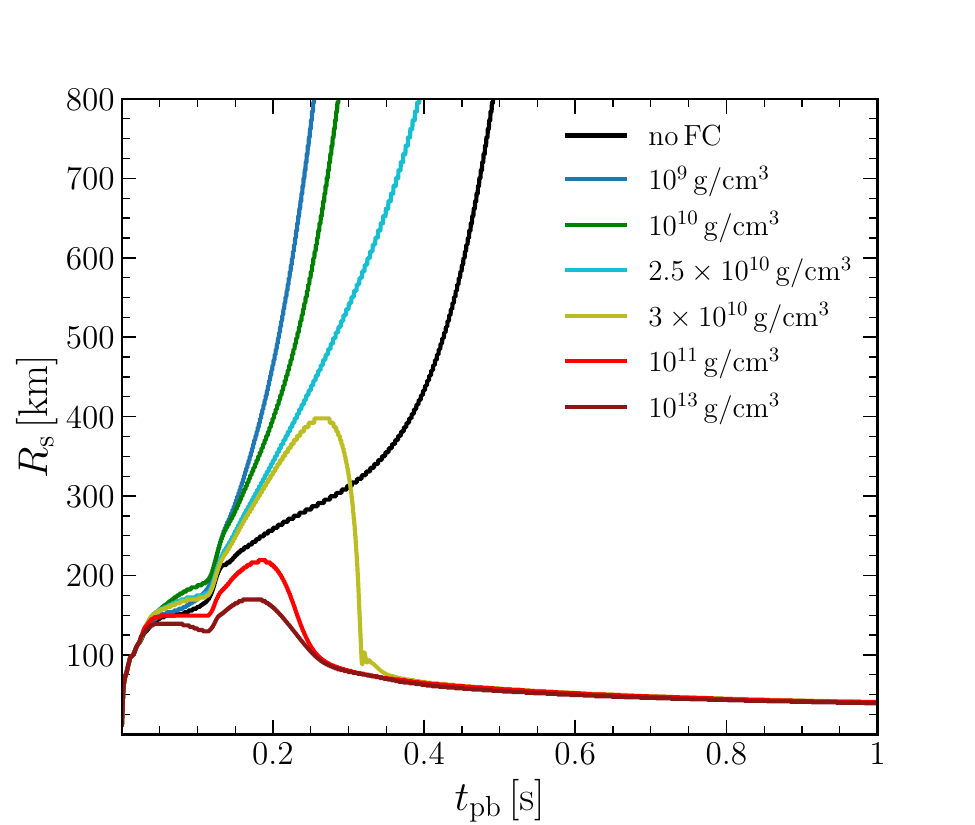}
    \caption{Shock radius evolution as a function of post-bounce time  for the  $16.5 \, M_\odot$ model with SFHo EOS. The black curve shows our reference CCSN model without flavor conversion (no FC). Colored curves correspond to CCSN models in which flavor equipartition is imposed below different $\rho_c$, ranging from $10^{9}$ to $10^{13}\, \mathrm{g/cm^3}$. 
    As $\rho_c$ increases, the CCSN models transition from successfully exploding to non-exploding.}
    \label{fig:16_5_shock_transition}
\end{figure}

Figure~\ref{fig:16_5_shock_transition} shows  the shock evolution for a range of $\rho_c$ values. 
As $\rho_c$ increases from $10^{9}$ to $10^{13}\,\mathrm{g/cm^3}$, the CCSN models transition from successfully exploding to non-exploding. This behavior highlights that the relative location of the spatial region affected by flavor conversion  with respect  to the gain region and the neutrinosphere radius is crucial in determining the fate of the CCSN explosion. 
If flavor conversion  occurs farther out, closer to the gain radius,  heating is enhanced; on the other hand, if flavor conversion occurs  deeper inside the proto-neutron star, it can weaken neutrino energy deposition  and hinder shock revival. 
These findings suggest that  flavor conversion can aid or hinder the explosion, confirming the early findings of  Refs.~\cite{Ehring2023abs,Akaho2026,Mori2025}. However, for a fixed  progenitor mass, the changes in energy deposition within the gain layer due to flavor conversion  can drastically modify the explosion fate. This transition is clearly illustrated for $\rho_c = 3 \times 10^{10}\,\mathrm{g/cm^3}$, where flavor conversion produces a distinct competing effect. Initially, flavor conversion within the gain region enhances the net heating and initiates shock expansion. However, because conversion also extends into the outer PNS cooling layer, converting electron flavors into  heavy-lepton neutrinos ($\nu_e, \bar{\nu}_e \to \nu_x$), 
thermal energy  leaks out of the core at an accelerated rate. This enhanced PNS cooling drains the core energy reservoir, ultimately suppressing the
electron-flavor luminosities. Consequently,  the gain-region heating rate becomes insufficient to sustain the shock expansion against the infalling accretion ram pressure, causing the shock to stall and recede.

\subsection{Sensitivity  to the equation of state}\label{sec:diff_EOS}
\begin{figure*}
    \centering
    \includegraphics[width=1\textwidth]{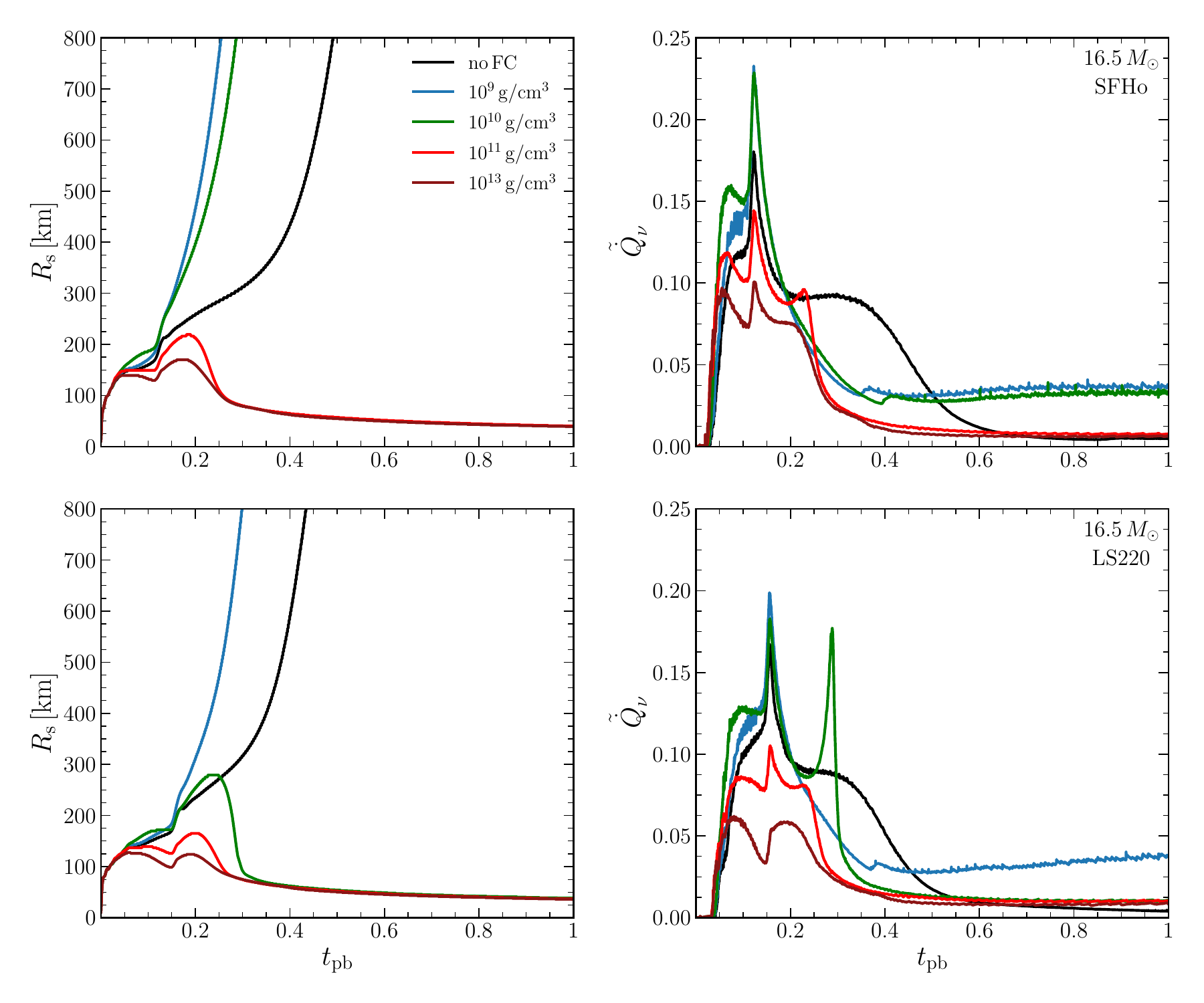}
    \caption{Shock radius (left panels) and dimensionless net neutrino heating rate $\tilde{\dot{Q}}_\nu$ (Eq.~\ref{eq:tildeQ}, right panels) as  functions of the post-bounce time  for the $16.5\,M_\odot$ progenitor with SFHo (top panels) and LS220 (bottom panels) EOSs. The colors of the curves follow the same convention as in Figs.~\ref{fig:4panel} and \ref{fig:16_5_shock_transition}. 
    The critical density $\rho_c$ at which the CCSN models  fail to explode depends on the EOS; while the SFHo model successfully explodes for $\rho_c = 10^{10}\, \mathrm{g/cm^{3}}$, the LS220 model does not. 
    }
    \label{fig:EOS}
\end{figure*}

The impact of flavor conversion on the CCSN explosion outcome depends  on the nuclear EOS. The two EOSs considered in this work, SFHo and LS220, are responsible for different  properties of dense nuclear matter~\cite{Oertel2017,Powell2025,Yasin2020,Schneider2019,Rusakov2026}. The SFHo EOS~\cite{STEINER2013} is derived within a relativistic mean-field framework calibrated to reproduce nuclear experimental constraints and neutron star observations, yielding a relatively soft EOS. In contrast, the LS220 EOS~\cite{LATTIMER1991} is based on a compressible liquid-drop model with a Skyrme-type interactions and a fixed incompressibility modulus of $220\, \mathrm{MeV}$, typically producing a stiffer behavior at subnuclear densities. 
According to the neutrino-driven explosion mechanism, the efficiency of shock revival depends on the balance between neutrino heating in the gain region and accretion ram pressure. The EOS plays a central role in this balance by setting the proto-neutron star contraction rate, the flavor-dependent neutrino luminosity and spectral energy distribution, as well as the properties  of the stalled shock. 

Figure~\ref{fig:EOS} compares the evolution of the shock radius and $\tilde{\dot{Q}}_\nu$ (Eq.~\ref{eq:tildeQ}) for the $16.5 \, M_\odot$ model with  SFHo (top panels) and LS220 (bottom panels) EOSs.
We find that the EOS significantly alters the neutrino heating conditions and, consequently, the shock evolution. 

The SFHo EOS leads to a larger gain region 
and  shallower density gradient. When flavor conversion is triggered below $\rho_c = 10^{10}\, \mathrm{g/cm^3}$, flavor conversion is responsible for a reduction in the average energy of $\nu_x$, indicating a net transfer of energy from $\nu_x$ to $\nu_e$ and $\bar{\nu}_e$. Since the electron flavors dominate charged-current absorption in the gain region,  the CCSN model exhibits a sustained increase in $\tilde{\dot{Q}}_\nu$, leading to a successful explosion. 

In contrast, the LS220 EOS is responsible for  a steeper density profile, and the shock  stalls at smaller radii. We find that the average $\nu_x$ energy increases due to flavor conversion, implying less energetic electron flavors. Moreover, because the shock radius is smaller and the density profile is steeper, the size of the gain region is reduced, leading to less efficient neutrino absorption.  
Consequently, the CCSN model with LS220 EOS and  flavor conversion below $\rho_c = 10^{10}\, \mathrm{g/cm^3}$ fails to revive the shock. A summary of the impact of the EOS on the explodability of the $16.5\, M_\odot$ model is  provided in Table~\ref{tab:runs}.

\subsection{Progenitor dependence}\label{sec:diff_progs}
\begin{figure*}
    \centering
    \includegraphics[width=0.95\textwidth]{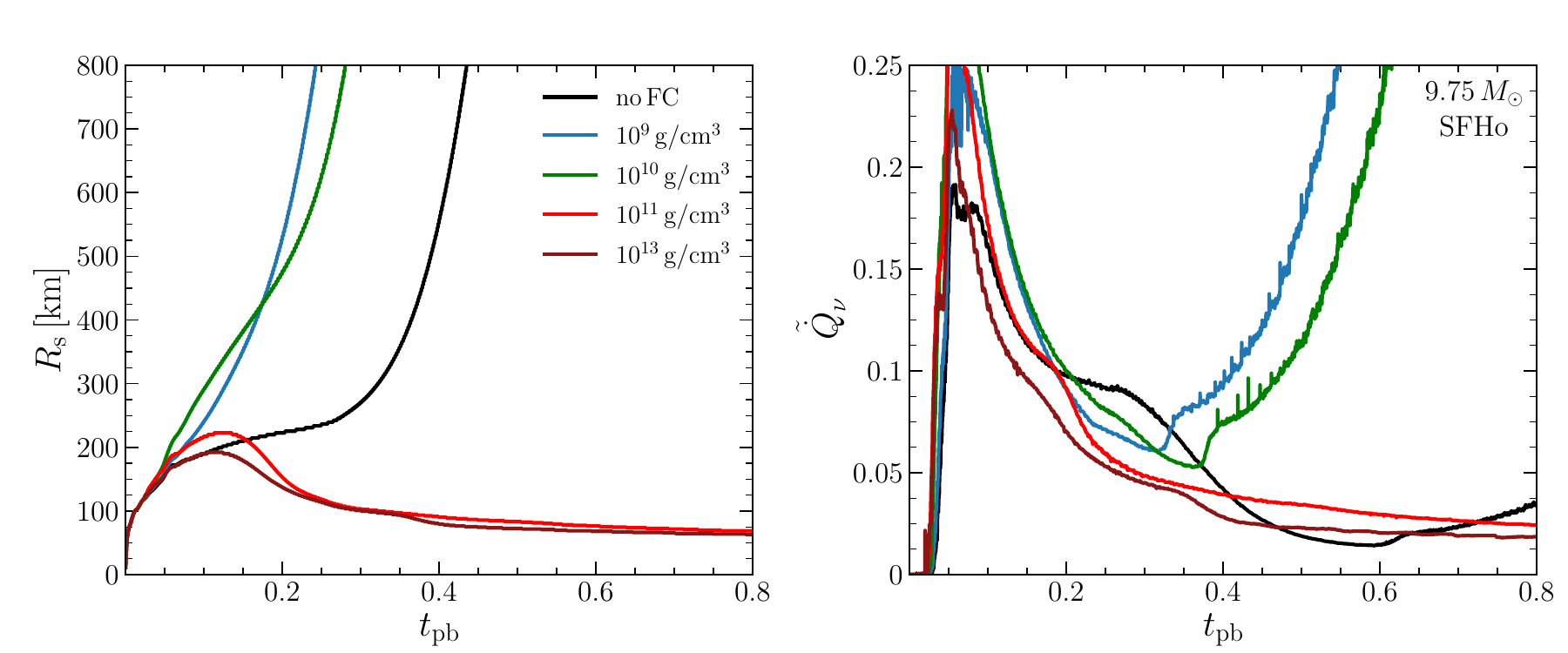}
    \includegraphics[width=0.95\textwidth]{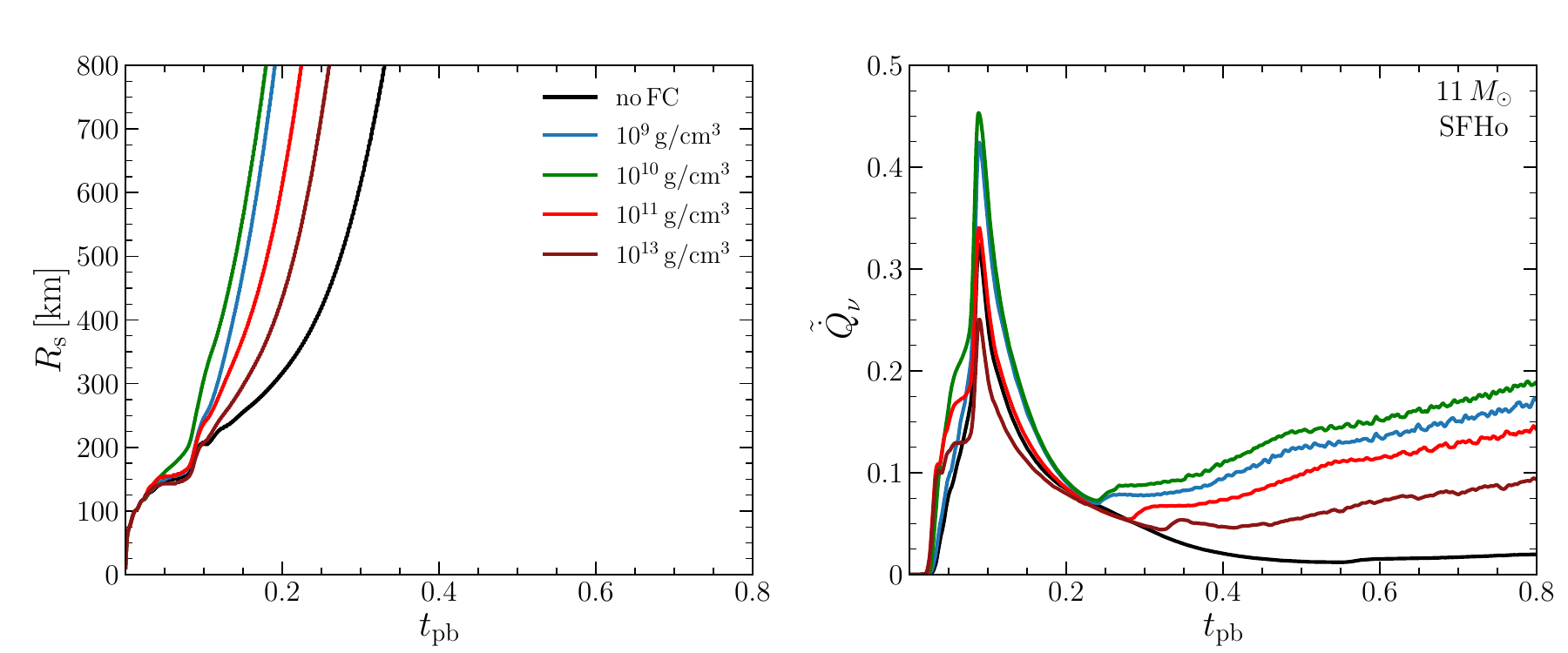}
    \caption{Explosion fate of low-mass CCSN models. Shock radius (left panel) and dimensionless net neutrino heating 
    (Eq.~\ref{eq:tildeQ}, right panel) as functions of the post-bounce 
    time. \textit{Top panels:} CCSN models with $9.75\,M_\odot$ ($\xi_{2.5} = 0.0001$) 
    and SFHo EOS. In the absence of flavor conversion and with   $\rho_c \lesssim 10^{10}\,\mathrm{g/cm^3}$, the CCSN models successfully explode, with flavor conversion leading to faster shock expansion and enhanced neutrino heating. However, for  $\rho_c \gtrsim 10^{11}\,\mathrm{g/cm^3}$ the models  fail to 
    explode. \textit{Bottom panels:} CCSN models with mass of $11\,M_\odot$  ($\xi_{2.5} = 0.01$) and SFHo EOS. All models successfully explode independent of $\rho_c$, with flavor conversion leading to faster shock expansion and enhanced neutrino heating.
    }   
    \label{fig:low_comp}
\end{figure*}

\begin{figure*}
    \includegraphics[width=0.95\textwidth]{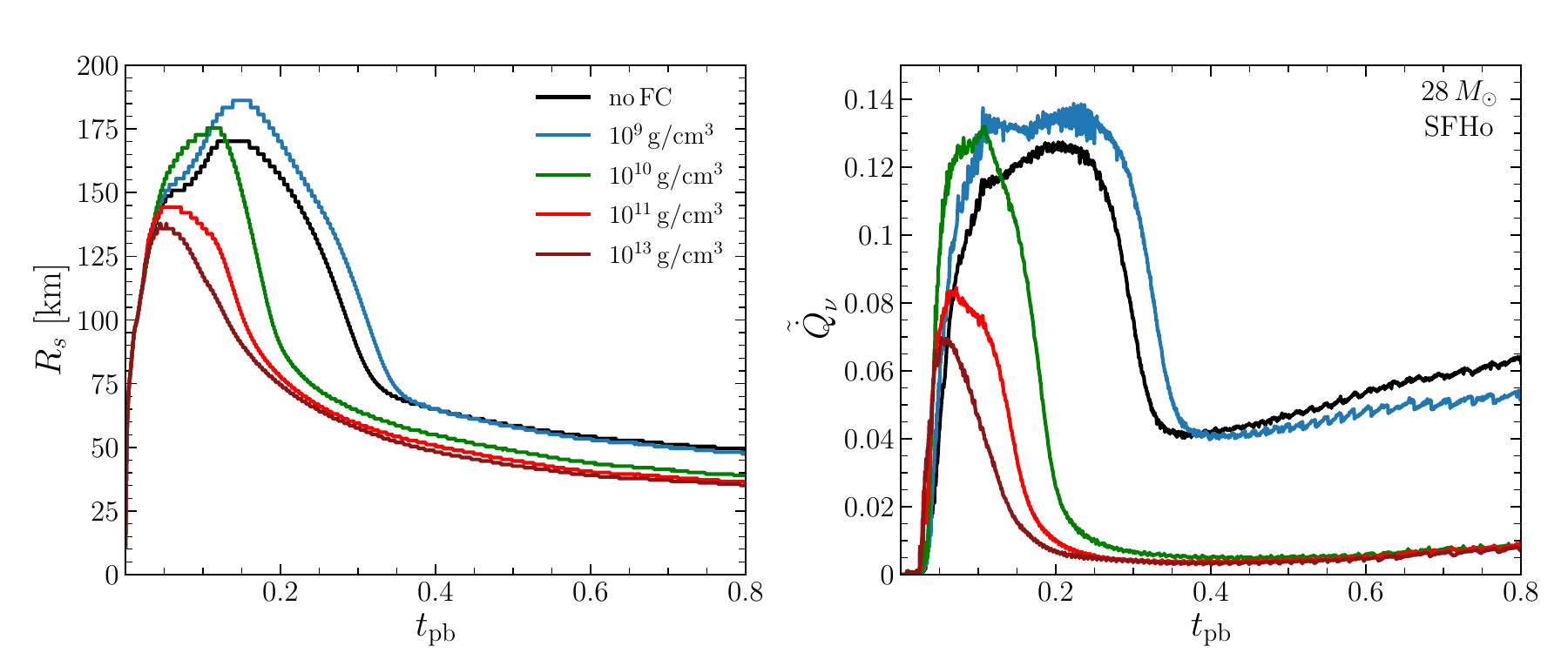}
    \includegraphics[width=0.95\textwidth]{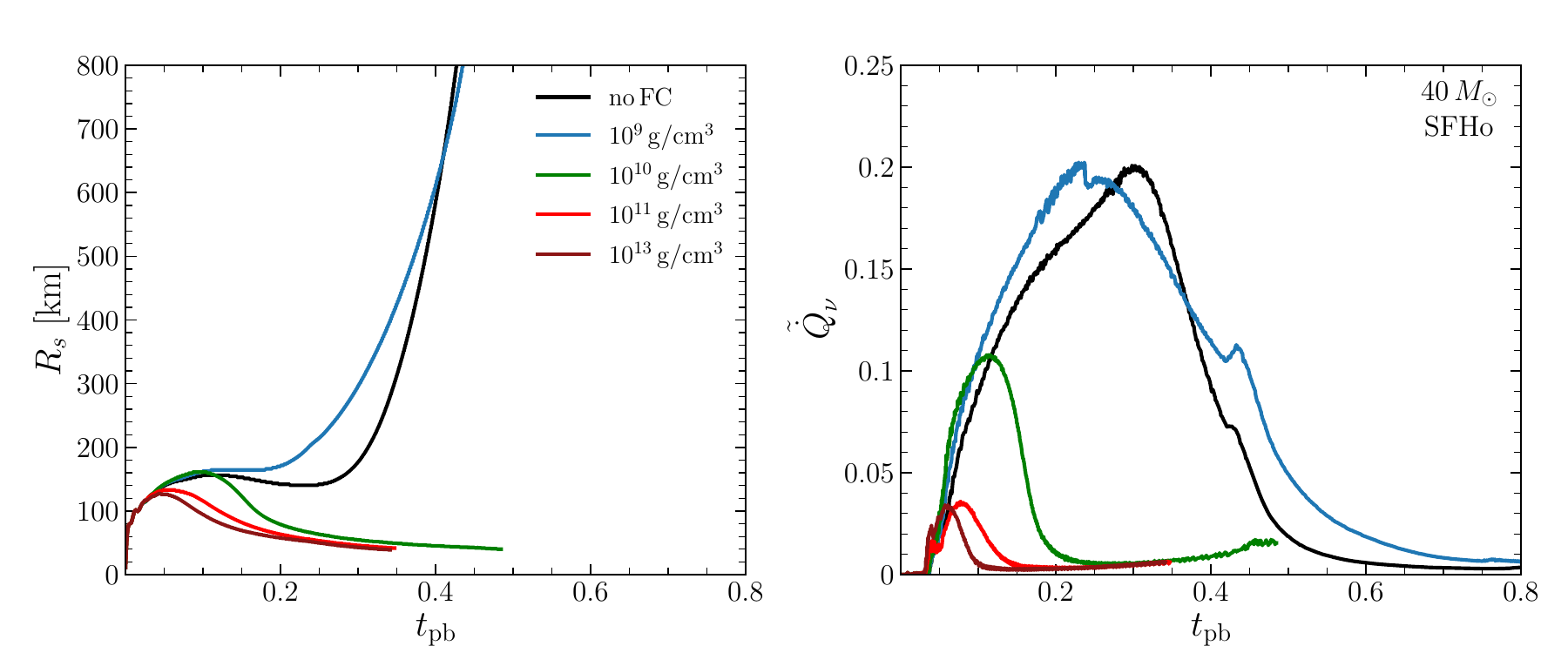}
    \includegraphics[width=0.95\textwidth]{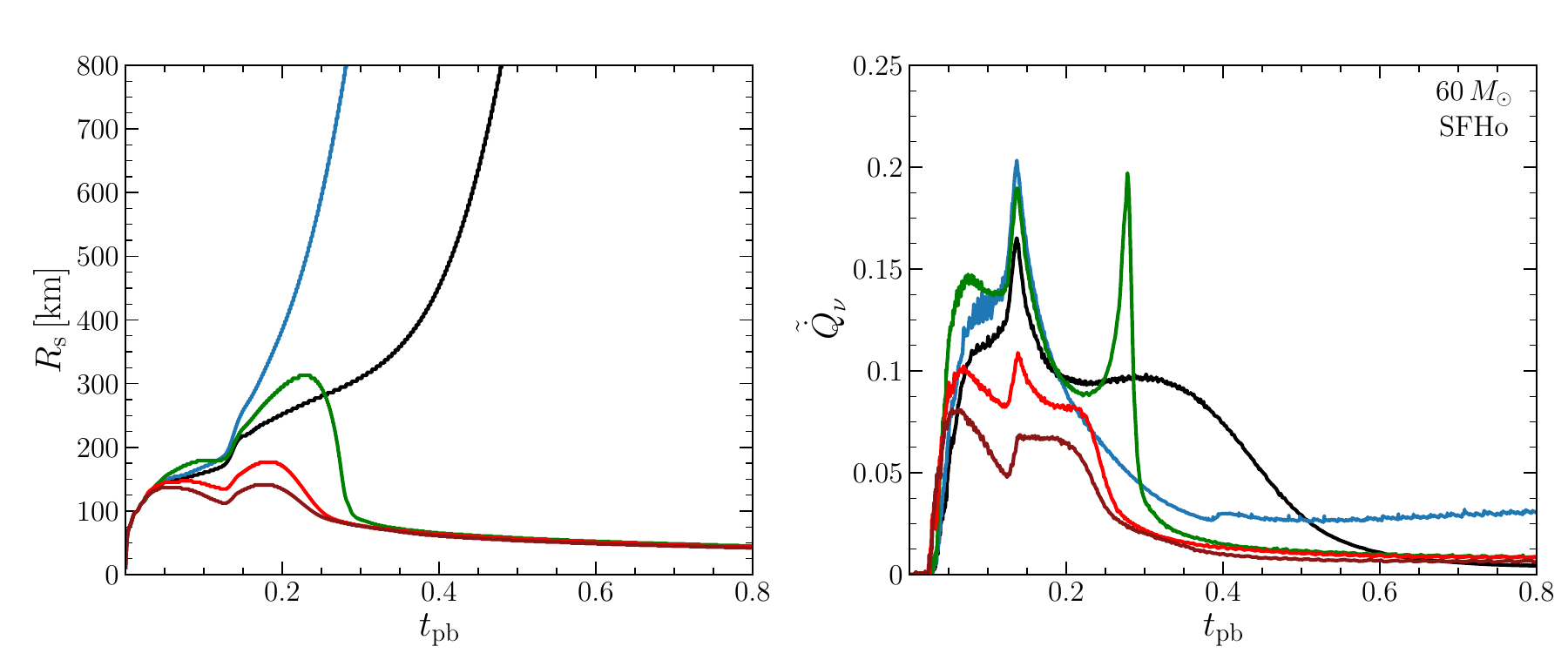}
    \caption{Same as Fig.~\ref{fig:low_comp}, but for one intermediate- and high-mass and compactness 
    progenitors  with SFHo EOS: $28\,M_\odot$ ($\xi_{2.5} = 0.28$, top panels), $40\,M_\odot$ ($\xi_{2.5} = 0.539$, middle panels), and $60\,M_\odot$ ($\xi_{2.5} = 0.174$, bottom panels). For the $28\,M_\odot$ model, 
    all models fail to explode regardless of $\rho_c$. For the high-compactness $40\,M_\odot$ model, only the model without flavor conversion and the one with  $\rho_c = 10^9\,\mathrm{g/cm^3}$ successfully explode, while the explosion is hindered for the models with $\rho_c \gtrsim 10^{10}\,\mathrm{g/cm^3}$. For the $60\,M_\odot$ progenitors, models without flavor conversion and with  $\rho_c \lesssim 10^{10}\,\mathrm{g/cm^3}$ successfully explode, while higher $\rho_c$ values quench the explosion. Hence, flavor conversion  changes the explosion fate depending on where it is triggered.}
    \label{fig:int_comp}
\end{figure*}

In order to investigate whether the impact of flavor conversion on the CCSN explosion depends on the progenitor structure,  we consider a set of progenitors with masses of $9.75\, M_\odot$, $11\, M_\odot$, $16.5\, M_\odot$, $28\, M_\odot$, $40\, M_\odot$, and $60\, M_\odot$, all simulated with SFHo EOS. We characterize each progenitor in terms of its compactness,
\begin{equation}
\label{eq:xi}
    \xi_M = \frac{M/M_\odot}{R(M)/1000\, \mathrm{km}} \, ,
\end{equation}
where $R(M)$ is the radial coordinate that encloses the mass $M$~\cite{OCONNOR2011}. The  compactness of  our models for $M=2.5\, M_\odot$ just before collapse  is provided in Table~\ref{tab:runs}. We note that our CCSN sample includes both low-compactness ($\xi_{2.5} \lesssim 0.45$) and high-compactness ($\xi_{2.5} \gtrsim 0.45$) progenitors, as well as low-, intermediate-, and high-mass stars, allowing us to probe the  parameter space of CCSN explodability.

In the following, we explore how the CCSN explodability and the mass of the remnant compact object ($M_{\mathrm{PNS}}$, we define the latter as the baryonic mass enclosed within the neutrinosphere averaged over $\nu_e$ and $\bar{\nu}_e$ and evaluated at $0.7\, \mathrm{s}$ after core bounce) change in the presence of flavor conversion, according to the spatial location where the latter is triggered.
\begin{table*}[t]
\centering
\caption{Summary of the impact of flavor conversion on our suite of CCSN models. For each model, we list the compactness parameter (cf.~Eq.~\ref{eq:xi}), the nuclear EOS, the critical density below which flavor conversion is triggered, the explosion outcome, and the remnant mass, from left to right respectively. Note that the remnant mass coincides with the baryonic neutron-star mass for exploding models, but should not be used as indicative of the mass of the black hole remnant for failed explosions.}
\begin{tabular*}{0.9\textwidth}{@{\extracolsep{\fill}}cccccc}
\toprule
Progenitor [$M_\odot$] & $\xi_{2.5}$ & EOS & $\rho_c$ [$\mathrm{g/cm^{3}}$] & Explosion & $M_\mathrm{PNS} \,[M_\odot]$ \\
\midrule
\multirow{5}{*}{$9.75$}
& \multirow{5}{*}{$0.0001$}
& \multirow{5}{*}{SFHo}
& - & Yes & $1.308$\\
& & & $10^9$ & Yes & $1.269$ \\
& & & $10^{10}$ & Yes & $1.277$\\
& & & $10^{11}$ & No & $1.324$\\
& & & $10^{13}$ & No  & $1.323$ \\
\midrule
\multirow{5}{*}{$11$}
& \multirow{5}{*}{$0.01$}
& \multirow{5}{*}{SFHo}
& - & Yes & $1.307$\\
& & & $10^9$ & Yes & $1.277$ \\
& & & $10^{10}$ & Yes & $1.273$\\
& & & $10^{11}$ & Yes & $1.284$\\
& & & $10^{13}$ & Yes  & $1.291$ \\
\midrule
\multirow{7}{*}{$16.5$}
& \multirow{7}{*}{$0.159$}
& \multirow{7}{*}{SFHo}
& - & Yes & $1.437$\\
& & & $10^9$ & Yes & $1.389$\\
& & & $10^{10}$ & Yes & $1.398$\\
& & & $2.5\times10^{10}$ & Yes & $1.422$\\
& & & $3\times10^{10}$ & No & $1.453$\\
& & & $10^{11}$ & No & $1.453$\\
& & & $10^{13}$ & No & $1.452$\\
\midrule
\multirow{5}{*}{$16.5$}
& \multirow{5}{*}{$0.159$}
& \multirow{5}{*}{LS220}
& - & Yes & $1.416$\\
& & & $10^9$ & Yes & $1.389$\\
& & & $10^{10}$ & No & $1.437$\\
& & & $10^{11}$ & No & $1.437$\\
& & & $10^{13}$ & No & $1.437$\\
\midrule
\multirow{5}{*}{$28$}
& \multirow{5}{*}{$0.28$}
& \multirow{5}{*}{SFHo}
& - & No & $1.632$\\
& & & $10^9$ & No & $1.632$\\
& & & $10^{10}$ & No & $1.632$\\
& & & $10^{11}$ & No & $1.631$\\
& & & $10^{13}$ & No & $1.630$\\
\midrule
\multirow{5}{*}{$40$}
& \multirow{5}{*}{$0.539$}
& \multirow{5}{*}{SFHo}
& - & Yes & $1.917$\\
& & & $10^9$ & Yes & $1.904$\\
& & & $10^{10}$ & No & $1.930$\\
& & & $10^{11}$ & No & $1.929$\\
& & & $10^{13}$ & No & $1.927$\\
\midrule
\multirow{5}{*}{$60$}
& \multirow{5}{*}{$0.174$}
& \multirow{5}{*}{SFHo}
& - & Yes & 1.499\\
& & & $10^9$ & Yes & $1.461$\\
& & & $10^{10}$ & No & $1.515$\\
& & & $10^{11}$ & No & $1.515$\\
& & & $10^{13}$ & No & $1.513$\\
\bottomrule
\end{tabular*}
\label{tab:runs}
\end{table*}

Figure~\ref{fig:low_comp} shows the shock radius and dimensionless net neutrino heating for two low-mass progenitors, both with low compactness: $9.75\,M_\odot$ ($\xi_{2.5} = 0.0001$) and 
$11\,M_\odot$ ($\xi_{2.5} = 0.01$). Both models explode without flavor conversion. For the $9.75\,M_\odot$ model, flavor conversion triggered at low $\rho_c$ ($\rho_c \lesssim 10^{10}\,\mathrm{g/cm^3}$)
enhances the explosion, while sufficiently high $\rho_c$ ($\rho_c \gtrsim 10^{11}\,\mathrm{g/cm^3}$) quenches it. These results  demonstrate that flavor conversion can have opposite effects on  the CCSN explodability for low-mass progenitors. For the $11\,M_\odot$ model, flavor conversion consistently enhances the explosion independent of $\rho_c$, 
leading to faster shock expansion and a less massive remnant, cf.~Table~\ref{tab:runs}. 
The qualitatively different response to flavor conversion between the two models can be understood by comparing their pre-explosion accretion histories. The $11\, M_\odot$ progenitor has a steeper density profile, leading to a more rapidly declining mass accretion rate after bounce. This causes the shock to expand faster already in the absence of flavor conversion, bringing the collapsing massive star  closer to the explosion threshold at earlier post-bounce times. In contrast, the $9.75\, M_\odot$ model sustains a higher accretion ram pressure for longer. Hence, the outcome is more sensitive to the magnitude of the heating modification: low $\rho_c$  (which enhances heating) tips the balance toward explosion, while high $\rho_c$  (which redistributes energy away from the gain region) is enough to suppress it.

The intermediate-mass, low-compactness $16.5\,M_\odot$ progenitor 
($\xi_{2.5} = 0.159$, Fig.~\ref{fig:4panel}) also supports the conjecture that flavor 
conversion can either enhance or quench the explosion depending on 
$\rho_c$, as discussed in Sec.~\ref{sec:diff_rhoc}.

Figure~\ref{fig:int_comp} shows the shock radius and dimensionless net neutrino heating for three progenitors: $28\,M_\odot$ ($\xi_{2.5} = 0.28$), $40\,M_\odot$ ($\xi_{2.5} = 0.539$), and 
$60\,M_\odot$ ($\xi_{2.5} = 0.174$), covering intermediate- and high-mass progenitors with both low and high compactness.
As for the intermediate-mass, $28\,M_\odot$ model ($\xi_{2.5} = 0.28$), independent of $\rho_c$, the CCSN models fail to explode, suggesting that flavor conversion is insufficient to overcome the ram pressure of the infalling material for this progenitor.

The $40\,M_\odot$ model is a high-mass, high-compactness progenitor ($\xi_{2.5} = 0.539$). In the absence of flavor conversion and  with flavor conversion triggered below $10^9\,\mathrm{g/cm^3}$, all models 
successfully explode. However, when  flavor conversion is triggered below $10^{10}\,\mathrm{g/cm^3}$, $10^{11}\,\mathrm{g/cm^3}$ and $10^{13}\,\mathrm{g/cm^3}$, the explosion is hindered. 

The $60\,M_\odot$ model is a high-mass, low-compactness progenitor ($\xi_{2.5} = 0.174$). The model without flavor conversion explodes, as do the models with flavor conversion triggered below $10^9\,\mathrm{g/cm^3}$ and $10^{10}\,\mathrm{g/cm^3}$, with the latter leading to enhanced neutrino heating and faster shock expansion. However, flavor conversion triggered at higher $\rho_c$ hinders the 
explosion. This model demonstrates that flavor conversion can  both favor or hinder the explosion of   high-mass progenitors.

Taken together, the results shown in 
Figs.~\ref{fig:low_comp} and \ref{fig:int_comp} suggest that flavor conversion affects  the fate of the core collapse depending on where it is triggered, across a wide range of progenitor masses and compactnesses.  
A summary of our findings is provided in Table~\ref{tab:runs}.

\section{\label{sec:D&C} Discussion and conclusions}
Neutrinos play a crucial role in the CCSN explosion mechanism. Although it is clear that flavor conversion of neutrinos occurs in the core of CCSNe, state-of-the-art neutrino-hydrodynamic  simulations do not account for this physics because of conceptual and computational limitations. 
In this work, for the first time, we explore the impact of flavor conversion on the CCSN explodability employing  models spanning a wide range of progenitor masses and two different nuclear EOSs (SFHo and LS220). We rely on  spherically symmetric models with mixing length treatment for convection and masses of $9.75\, M_\odot$, $11\, M_\odot$,  $16.5\, M_\odot$,  $28\, M_\odot$,  $40\, M_\odot$, and  $60\, M_\odot$.
Because of the uncertainties linked to our understanding of flavor conversion and the technical challenges intrinsic to the embedment of neutrino kinetics in CCSN simulations, we model flavor conversion through  a schematic approach and impose that flavor equipartition is achieved instantaneously below a critical baryon density ($\rho_c$),  that we vary  between the neutrinosphere and the shock radius.
This  setup allows us to probe how the region of the CCSN core where flavor conversion occurs affects shock revival and the resulting  properties of the remnant compact object.

We find that flavor conversion can change the core-collapse fate  regardless of the progenitor compactness, progenitor mass, or EOS, depending on the spatial region where it is  triggered. The nuclear EOS further affects the  threshold value of $\rho_c$ at which the CCSN models transition from exploding to non-exploding or vice versa.  

For the 
$11\,M_\odot$ model ($\xi_{2.5} = 0.01$), which successfully explodes 
in the absence of flavor conversion, the latter consistently enhances 
neutrino heating independent of $\rho_c$, accelerating shock expansion 
and lowering the remnant mass due to reduced accretion. For the 
$9.75\,M_\odot$ model, flavor conversion triggered at low $\rho_c$ ($\rho_c \lesssim 10^{10}\,\mathrm{g/cm^3}$) 
enhances the explosion, while sufficiently high $\rho_c$ ($\rho_c \gtrsim 10^{11}\,\mathrm{g/cm^3}$) quenches it. 
In contrast to the early findings of 
Refs.~\cite{Ehring2023abs, Mori2025, Akaho2026}, suggesting that 
flavor transformation  aids the CCSN explodability of low-mass 
progenitors, we find that the impact of flavor conversion on the explodability of low-mass progenitors is nuanced and depends on the region  where flavor conversion is triggered.

For intermediate-mass progenitors, we find a similar trend,  regardless of their compactness. 
The 
$16.5\,M_\odot$ model ($\xi_{2.5} = 0.159$), which explodes without 
flavor conversion, fails to explode when flavor conversion is triggered 
at high densities ($\rho_c \gtrsim 10^{10}\,\mathrm{g/cm^3}$), while 
low $\rho_c$ further enhances the explosion. The $28\,M_\odot$ model 
($\xi_{2.5} = 0.222$) fails to explode regardless of $\rho_c$.

As for high-mass progenitors, 
%the explosion outcome again depends 
%critically on $\rho_c$ regardless of the compactness. 
the high-compactness 
$40\,M_\odot$ model ($\xi_{2.5} = 0.54$) explodes without flavor 
conversion and with flavor conversion triggered below 
$10^9\,\mathrm{g/cm^3}$, but fails when flavor conversion is triggered 
at higher $\rho_c$. The high-mass, low-compactness $60\,M_\odot$ model 
($\xi_{2.5} = 0.174$) explodes without flavor conversion and with 
 $\rho_c \lesssim 10^{10}\,\mathrm{g/cm^3}$, but fails at higher $\rho_c$. In contrast 
to the early findings of Refs.~\cite{Ehring2023abs, Akaho2026}, which 
concluded that flavor conversion is  insufficient to overcome the 
ram pressure of infalling material in CCSNe with high accretion rate, 
we find that the explodability of  high-mass progenitors can be enhanced or hindered by flavor 
conversion.

Our findings show  that flavor conversion can change 
the CCSN fate depending on where it is triggered, regardless of 
the progenitor mass, compactness, or nuclear EOS.  This overall behavior reflects a 
delicate interplay between neutrino heating and the ram pressure of the 
accreting material. Correspondingly, the rate of failed explosions is affected by flavor conversion as explored in Ref.~\cite{Gogilashvili2026_prl}. The compact remnant mass is also strongly 
affected~\cite{Gogilashvili2026_prl}: enhanced explosions lead to lower remnant masses, whereas 
failed explosions result in slightly higher remnant masses due to prolonged 
accretion.

The properties of the Si/O interface are crucial to further understand 
the trend reported above as a function of core compactness. In fact, 
the accretion of infalling matter onto this interface leads to a sudden 
drop in the mass accretion rate and ram pressure at the shock, which 
may facilitate shock 
revival~\cite{Wang2022_prog_study_ram_pressure, Boccioli2023_explodability, Boccioli2025}.
For instance, in our $11\,M_\odot$ model, the accretion of the Si/O 
interface onto the shock at $\sim 0.1\,\mathrm{s}$ 
(cf.~Fig.~\ref{fig:low_comp}) produces a sharp drop in the ram 
pressure. Such drop in the mass accretion rate, combined with enhanced 
heating from flavor conversion, results in a rapid explosion. In 
contrast, for high-mass progenitors, such as $60\,M_\odot$ models, the 
Si/O interface typically reaches the shock at 
$\sim 0.15\text{--}0.2\,\mathrm{s}$ (Fig.~\ref{fig:int_comp}). If 
flavor transformation enhances neutrino heating (e.g., $\rho_c = 
10^9\,\mathrm{g/cm^3}$) when the Si/O interface is accreted, the 
explosion can be triggered or accelerated. Conversely, when flavor 
conversion suppresses heating (e.g., $\rho_c = 
10^{11}\,\mathrm{g/cm^3}$), the beneficial impact of the Si/O 
interface on the explosion is hindered. Hence, flavor transformation 
can either amplify or counteract the effect of the Si/O interface, 
further contributing to the non-trivial dependence of the CCSN 
explodability on the progenitor structure and the region of the CCSN 
core affected by flavor conversion.

We also note that, in our exploding models, the shock does not exhibit a stalled phase, instead undergoes relatively prompt expansion once favorable heating conditions are achieved. Although our 1D+ framework does not fully capture multidimensional effects, this behavior suggests that  the development of large-scale hydrodynamic instabilities, such as the standing accretion shock instability~\cite{Blondin2003,Blondin2007,Foglizzo2012,Fernandez2010}, might be disfavored.

Finally, we emphasize that flavor conversion in this work is implemented through a  parametric prescription. Although this approach allows us to isolate and understand the macroscopic impact of neutrinos on the CCSN physics, the coupling of neutrino kinetics to the CCSN  hydrodynamics is crucial to assess the implications of neutrino physics on the CCSN explosion dynamics and the multi-messenger observables. Moreover, the consequences of flavor conversion on the CCSN explodability should be considered together with other ingredients potentially important for the explosion mechanism, such as magnetic field amplification~\cite{Obergaulinger2014,Mueller2020}, muon production in the newly-born neutron star~\cite{Bollig2017},   large-scale perturbations from convective  oxygen and silicon shell burning~\cite{Mueller2017},  corrections to  neutral current neutrino-nucleon scattering at low densities~\cite{Horowitz2017},  or the effective mass of nucleons  above nuclear saturation density~\cite{Yasin2020}. 
Future work in this direction will be essential to 
assess the role of flavor transformation in shaping the properties of the remnant compact object.

\section*{Acknowledgments}
We thank Luca Boccioli, Thomas Janka and Evan P.~O'Connor  for insightful discussions.
This project has received support from the European Union (ERC, ANET, Project No.~101087058) and the Villum Foundation (Project No.~13164).
Views and opinions expressed are those of the authors only and do not necessarily reflect those of the European Union or the European Research Council. Neither the European Union nor the granting authority can be held responsible for them. The Tycho supercomputer hosted at the SCIENCE HPC center at the University of Copenhagen was used to perform the numerical simulations whose results are presented in this paper.

\bibliography{References}

@article{low2004,
  title = {Control of star formation by supersonic turbulence},
  author = {Mac Low, Mordecai-Mark and Klessen, Ralf S.},
  journal = {Rev. Mod. Phys.},
  volume = {76},
  issue = {1},
  pages = {125--194},
  numpages = {0},
  year = {2004},
  month = {Jan},
  publisher = {American Physical Society},
  doi = {10.1103/RevModPhys.76.125},
  url ={https://link.aps.org/doi/10.1103/RevModPhys.76.125}
}

@ARTICLE{woosley2002,
       author = {{Woosley}, S.~E. and {Heger}, A. and {Weaver}, T.~A.},
        title = "{The evolution and explosion of massive stars}",
      journal = {Rev. Mod. Phys.},
         year = 2002,
        month = nov,
       volume = {74},
       number = {4},
        pages = {1015-1071},
          doi = {10.1103/RevModPhys.74.1015},
       url = {https://ui.adsabs.harvard.edu/abs/2002RvMP...74.1015W}
}

@ARTICLE{DIEHL2021,
author = "Diehl, Roland and others",
    title = "{The radioactive nuclei and in the Cosmos and in the solar system}",
    eprint = "2109.08558",
    archivePrefix = "arXiv",
    primaryClass = "astro-ph.HE",
    doi = "10.1017/pasa.2021.48",
    journal = "Publ. Astron. Soc. Austral.",
    volume = "38",
    pages = "e062",
    year = "2021"
    }

@ARTICLE{HILLEBRANDT1981,
       author = {{Hillebrandt}, W. and {M\"ueller}, E.},
        title = "{Computer simulations of stellar collapse and shock wave propagation}",
      journal = {Astron. Astrophys.},
         year = 1981,
        month = nov,
       volume = {103},
       number = {1},
        pages = {147-153},
       url = {https://ui.adsabs.harvard.edu/abs/1981A&A...103..147H}
}

@ARTICLE{MAZUREK1982,
       author = {{Mazurek}, T.~J.},
        title = "{The energetics of adiabatic shocks in stellar collapse}",
      journal = {Astrophys. J. Lett.},
         year = 1982,
        month = aug,
       volume = {259},
        pages = {L13-L17},
          doi = {10.1086/183839},
       url = {https://ui.adsabs.harvard.edu/abs/1982ApJ...259L..13M}
}

@INPROCEEDINGS{MAZUREK82,
       author = {{Mazurek}, T.~J. and {Cooperstein}, J. and {Kahana}, S.},
        title = "{Shock stagnation and neutrino losses in stellar collapse}",
    booktitle = {Supernovae: A Survey of Current Research},
         year = 1982,
       editor = {{Rees}, M.~J. and {Stoneham}, R.~J.},
       series = {NATO Advanced Study Institute (ASI) Series C},
       volume = {90},
        month = nov,
        pages = {71-77},
       url = {https://ui.adsabs.harvard.edu/abs/1982ASIC...90...71M}
}

@ARTICLE{Burrows2007,
       author = {{Burrows}, Adam and {Dessart}, Luc and {Ott}, Christian D. and {Livne}, Eli},
        title = "{Multi-dimensional explorations in supernova theory}",
      journal = {Phys. Rept.},
         year = 2007,
        month = apr,
       volume = {442},
       number = {1-6},
        pages = {23-37},
          doi = {10.1016/j.physrep.2007.02.001},
archivePrefix = {arXiv},
       eprint = {astro-ph/0612460},
 primaryClass = {astro-ph},
       adsurl = {https://ui.adsabs.harvard.edu/abs/2007PhR...442...23B}
}

@ARTICLE{Rusakov2026,
 author = "Rusakov, Aleksandr and Burrows, Adam S. and Wang, Tianshu and Vartanyan, David",
    title = "{An Exploration of the Equation of State Dependence of Core-Collapse Supernova Explosion Outcomes and Signatures}",
    journal = {arXiv e-prints},
    eprint = "2602.09025",
    archivePrefix = "arXiv",
    primaryClass = "astro-ph.HE",
    month = "2",
    year = "2026"
}

@ARTICLE{BURROWS1986,
       author = {{Burrows}, A. and {Lattimer}, J.~M.},
        title = "{The Birth of Neutron Stars}",
      journal = {Astrophys. J.},
         year = 1986,
        month = aug,
       volume = {307},
        pages = {178},
          doi = {10.1086/164405},
       url = {https://ui.adsabs.harvard.edu/abs/1986ApJ...307..178B}
}

@ARTICLE{Burrows2025,
       author = {{Burrows}, Adam and {Wang}, Tianshu and {Vartanyan}, David},
        title = "{Channels of Stellar-mass Black Hole Formation}",
      journal = {\apj},
         year = 2025,
        month = jul,
       volume = {987},
       number = {2},
          eid = {164},
        pages = {164},
          doi = {10.3847/1538-4357/addd04},
archivePrefix = {arXiv},
       eprint = {2412.07831},
 primaryClass = {astro-ph.SR},
       adsurl = {https://ui.adsabs.harvard.edu/abs/2025ApJ...987..164B}
}

@ARTICLE{Bethe1990RvMP,
       author = {{Bethe}, H.~A.},
        title = "{Supernova mechanisms}",
      journal = {Rev.  Mod. Phys.},
         year = 1990,
        month = oct,
       volume = {62},
       number = {4},
        pages = {801-866},
          doi = {10.1103/RevModPhys.62.801},
       adsurl = {https://ui.adsabs.harvard.edu/abs/1990RvMP...62..801B}
}

@ARTICLE{FISCHER2009,
       author = {{Fischer}, T. and {Whitehouse}, S.~C. and {Mezzacappa}, A. and {Thielemann}, F. -K. and {Liebend{\"o}rfer}, M.},
        title = "{The neutrino signal from protoneutron star accretion and black hole formation}",
      journal = {Astron. Astrophys.},
         year = 2009,
        month = may,
       volume = {499},
       number = {1},
        pages = {1-15},
          doi = {10.1051/0004-6361/200811055},
archivePrefix = {arXiv},
       eprint = {0809.5129},
 primaryClass = {astro-ph},
       url = {https://ui.adsabs.harvard.edu/abs/2009A&A...499....1F}
}

@ARTICLE{OCONNOR2011,
       author = {{O'Connor}, Evan and {Ott}, Christian D.},
        title = "{Black Hole Formation in Failing Core-Collapse Supernovae}",
      journal = {Astrophys. J.},
         year = 2011,
        month = apr,
       volume = {730},
       number = {2},
          eid = {70},
        pages = {70},
          doi = {10.1088/0004-637X/730/2/70},
archivePrefix = {arXiv},
       eprint = {1010.5550},
 primaryClass = {astro-ph.HE},
       url = {https://ui.adsabs.harvard.edu/abs/2011ApJ...730...70O}
}

@ARTICLE{COLGATE1966,
       author = {{Colgate}, Stirling A. and {White}, Richard H.},
        title = "{The Hydrodynamic Behavior of Supernovae Explosions}",
      journal = {Astrophys. J.},
         year = 1966,
        month = mar,
       volume = {143},
        pages = {626},
          doi = {10.1086/148549},
       url = {https://ui.adsabs.harvard.edu/abs/1966ApJ...143..626C}
}

@ARTICLE{BETHE1985,
       author = {{Bethe}, H.~A. and {Wilson}, J.~R.},
        title = "{Revival of a stalled supernova shock by neutrino heating}",
      journal = {Astrophys. J.},
         year = 1985,
        month = aug,
       volume = {295},
        pages = {14-23},
          doi = {10.1086/163343},
       url = {https://ui.adsabs.harvard.edu/abs/1985ApJ...295...14B}
}

@article{STEINER2013,
 author = "Steiner, Andrew W. and Hempel, Matthias and Fischer, Tobias",
    title = "{Core-collapse supernova equations of state based on neutron star observations}",
    eprint = "1207.2184",
    archivePrefix = "arXiv",
    primaryClass = "astro-ph.SR",
    reportNumber = "INT-PUB-12-033",
    doi = "10.1088/0004-637X/774/1/17",
    journal = "Astrophys. J.",
    volume = "774",
    pages = "17",
    year = "2013"
}

@ARTICLE{JANKA2012,
       author = {{Janka}, Hans-Thomas},
        title = "{Explosion Mechanisms of Core-Collapse Supernovae}",
      journal = {Ann. Rev. Nucl. Part. Sci.},
         year = 2012,
        month = nov,
       volume = {62},
       number = {1},
        pages = {407-451},
          doi = {10.1146/annurev-nucl-102711-094901},
archivePrefix = {arXiv},
       eprint = {1206.2503},
 primaryClass = {astro-ph.SR},
       url = {https://ui.adsabs.harvard.edu/abs/2012ARNPS..62..407J}
}

@ARTICLE{Duan2010,
       author = {{Duan}, Huaiyu and {Fuller}, George M. and {Qian}, Yong-Zhong},
        title = "{Collective Neutrino Oscillations}",
      journal = {Ann. Rev. Nucl. Part. Sci.},
         year = 2010,
        month = nov,
       volume = {60},
        pages = {569-594},
          doi = {10.1146/annurev.nucl.012809.104524},
archivePrefix = {arXiv},
       eprint = {1001.2799},
 primaryClass = {hep-ph},
       url = {https://ui.adsabs.harvard.edu/abs/2010ARNPS..60..569D}
}

@ARTICLE{Mirizzi2016,
       author = {{Mirizzi}, A. and {Tamborra}, I. and {Janka}, H.-T. and {Saviano}, N. and {Scholberg}, K. and {Bollig}, R. and {H{\"u}depohl}, L. and {Chakraborty}, S.},
        title = "{Supernova neutrinos: production, oscillations and detection}",
      journal = {Riv. Nuovo Cim.},
         year = 2016,
        month = feb,
       volume = {39},
       number = {1-2},
        pages = {1-112},
          doi = {10.1393/ncr/i2016-10120-8},
archivePrefix = {arXiv},
       eprint = {1508.00785},
 primaryClass = {astro-ph.HE},
       url = {https://ui.adsabs.harvard.edu/abs/2016NCimR..39....1M}
}

@ARTICLE{Sawyer2005,
       author = {{Sawyer}, R.~F.},
        title = "{Speed-up of neutrino transformations in a supernova environment}",
      journal = {Phys. Rev. D},
         year = 2005,
        month = aug,
       volume = {72},
       number = {4},
          eid = {045003},
        pages = {045003},
          doi = {10.1103/PhysRevD.72.045003},
archivePrefix = {arXiv},
       eprint = {hep-ph/0503013},
 primaryClass = {astro-ph},
       url = {https://ui.adsabs.harvard.edu/abs/2005PhRvD..72d5003S}
}

@ARTICLE{Sawyer2016,
       author = {{Sawyer}, R.~F.},
        title = "{Neutrino Cloud Instabilities Just above the Neutrino Sphere of a Supernova}",
      journal = {Phys. Rev. Lett.},
         year = 2016,
        month = feb,
       volume = {116},
       number = {8},
          eid = {081101},
        pages = {081101},
          doi = {10.1103/PhysRevLett.116.081101},
archivePrefix = {arXiv},
       eprint = {1509.03323},
 primaryClass = {astro-ph.HE},
       url = {https://ui.adsabs.harvard.edu/abs/2016PhRvL.116h1101S}
}

@article{Ehring2023,
  title = {Fast neutrino flavor conversion in core-collapse supernovae: A parametric study in 1D models},
  author = {Ehring, Jakob and Abbar, Sajad and Janka, Hans-Thomas and Raffelt, Georg G. and Tamborra, Irene},
  journal = {Phys. Rev. D},
  volume = {107},
  issue = {10},
  pages = {103034},
  numpages = {20},
  year = {2023},
  month = {May},
  publisher = {American Physical Society},
  doi = {10.1103/PhysRevD.107.103034},
  url = {https://link.aps.org/doi/10.1103/PhysRevD.107.103034}
}

@ARTICLE{Blondin2003,
       author = {{Blondin}, John M. and {Mezzacappa}, Anthony and {DeMarino}, Christine},
        title = "{Stability of Standing Accretion Shocks, with an Eye toward Core-Collapse Supernovae}",
      journal = {\apj},
         year = 2003,
        month = feb,
       volume = {584},
       number = {2},
        pages = {971-980},
          doi = {10.1086/345812},
archivePrefix = {arXiv},
       eprint = {astro-ph/0210634},
 primaryClass = {astro-ph},
       adsurl = {https://ui.adsabs.harvard.edu/abs/2003ApJ...584..971B}
}

@ARTICLE{Blondin2007,
  author = "Blondin, John M. and Mezzacappa, Anthony",
    title = "{Pulsar spins from an instability in the accretion shock of supernovae}",
    eprint = "astro-ph/0611680",
    archivePrefix = "arXiv",
    doi = "10.1038/nature05428",
    journal = "Nature",
    volume = "445",
    pages = "58",
    year = "2007"
}

@ARTICLE{Foglizzo2012,
       author = {{Foglizzo}, Thierry and {Masset}, Fr{\'e}d{\'e}ric and {Guilet}, J{\'e}r{\^o}me and {Durand}, Gilles},
        title = "{Shallow Water Analogue of the Standing Accretion Shock Instability: Experimental Demonstration and a Two-Dimensional Model}",
      journal = {\prl},
         year = 2012,
        month = feb,
       volume = {108},
       number = {5},
          eid = {051103},
        pages = {051103},
          doi = {10.1103/PhysRevLett.108.051103},
archivePrefix = {arXiv},
       eprint = {1112.3448},
 primaryClass = {astro-ph.HE},
       adsurl = {https://ui.adsabs.harvard.edu/abs/2012PhRvL.108e1103F}
}

@ARTICLE{Fernandez2010,
       author = {{Fern{\'a}ndez}, Rodrigo},
        title = "{The Spiral Modes of the Standing Accretion Shock Instability}",
      journal = {\apj},
         year = 2010,
        month = dec,
       volume = {725},
       number = {2},
        pages = {1563-1580},
          doi = {10.1088/0004-637X/725/2/1563},
archivePrefix = {arXiv},
       eprint = {1003.1730},
 primaryClass = {astro-ph.SR},
       adsurl = {https://ui.adsabs.harvard.edu/abs/2010ApJ...725.1563F}
}

@ARTICLE{Mueller2017,
       author = {{M{\"u}ller}, Bernhard and {Melson}, Tobias and {Heger}, Alexander and {Janka}, Hans-Thomas},
        title = "{Supernova simulations from a 3D progenitor model - Impact of perturbations and evolution of explosion properties}",
      journal = {Mon. Not. Roy. Astron. Soc.},
         year = 2017,
        month = nov,
       volume = {472},
       number = {1},
        pages = {491-513},
          doi = {10.1093/mnras/stx1962},
archivePrefix = {arXiv},
       eprint = {1705.00620},
 primaryClass = {astro-ph.SR},
       adsurl = {https://ui.adsabs.harvard.edu/abs/2017MNRAS.472..491M}
}

@ARTICLE{Powell2025,
 author = {Powell, Jade and M{\"u}ller, Bernhard},
    title = "{Impact of the nuclear equation of state on the explodability of massive stars}",
    journal = {arXiv e-prints},
    eprint = "2510.20076",
    archivePrefix = "arXiv",
    primaryClass = "astro-ph.HE",
    month = "10",
    year = "2025"
}

@ARTICLE{Bollig2017,
       author = {{Bollig}, R. and {Janka}, H.-T. and {Lohs}, A. and {Mart{\'\i}nez-Pinedo}, G. and {Horowitz}, C.~J. and {Melson}, T.},
        title = "{Muon Creation in Supernova Matter Facilitates Neutrino-Driven Explosions}",
      journal = {Phys. Rev. Lett.},
         year = 2017,
        month = dec,
       volume = {119},
       number = {24},
          eid = {242702},
        pages = {242702},
          doi = {10.1103/PhysRevLett.119.242702},
archivePrefix = {arXiv},
       eprint = {1706.04630},
 primaryClass = {astro-ph.HE},
       adsurl = {https://ui.adsabs.harvard.edu/abs/2017PhRvL.119x2702B}
}

@ARTICLE{Horowitz2017,
       author = {{Horowitz}, C.~J. and {Caballero}, O.~L. and {Lin}, Zidu and {O'Connor}, Evan and {Schwenk}, A.},
        title = "{Neutrino-nucleon scattering in supernova matter from the virial expansion}",
      journal = {\prc},
         year = 2017,
        month = feb,
       volume = {95},
       number = {2},
          eid = {025801},
        pages = {025801},
          doi = {10.1103/PhysRevC.95.025801},
archivePrefix = {arXiv},
       eprint = {1611.05140},
 primaryClass = {nucl-th},
       adsurl = {https://ui.adsabs.harvard.edu/abs/2017PhRvC..95b5801H}
}

@ARTICLE{Yasin2020,
       author = {{Yasin}, H. and {Sch{\"a}fer}, S. and {Arcones}, A. and {Schwenk}, A.},
        title = "{Equation of State Effects in Core-Collapse Supernovae}",
      journal = {\prl},
         year = 2020,
        month = mar,
       volume = {124},
       number = {9},
          eid = {092701},
        pages = {092701},
          doi = {10.1103/PhysRevLett.124.092701},
archivePrefix = {arXiv},
       eprint = {1812.02002},
 primaryClass = {nucl-th},
       adsurl = {https://ui.adsabs.harvard.edu/abs/2020PhRvL.124i2701Y}
}

@ARTICLE{Schneider2019,
       author = {{Schneider}, A.~S. and {Roberts}, L.~F. and {Ott}, C.~D. and {O'Connor}, E.},
        title = "{Equation of state effects in the core collapse of a 20 -M$_{\odot}$ star}",
      journal = {\prc},
         year = 2019,
        month = nov,
       volume = {100},
       number = {5},
          eid = {055802},
        pages = {055802},
          doi = {10.1103/PhysRevC.100.055802},
archivePrefix = {arXiv},
       eprint = {1906.02009},
 primaryClass = {astro-ph.HE},
       adsurl = {https://ui.adsabs.harvard.edu/abs/2019PhRvC.100e5802S}
}

@ARTICLE{Mueller2020,
       author = {{M{\"u}ller}, Bernhard and {Varma}, Vishnu},
        title = "{A 3D simulation of a neutrino-driven supernova explosion aided by convection and magnetic fields}",
      journal = {Mon. Not. Roy. Astron. Soc.},
         year = 2020,
        month = nov,
       volume = {498},
       number = {1},
        pages = {L109-L113},
          doi = {10.1093/mnrasl/slaa137},
archivePrefix = {arXiv},
       eprint = {2007.04775},
 primaryClass = {astro-ph.HE},
       adsurl = {https://ui.adsabs.harvard.edu/abs/2020MNRAS.498L.109M}
}

@ARTICLE{Obergaulinger2014,
       author = {{Obergaulinger}, M. and {Janka}, H.-Th. and {Aloy}, M.~{\'A}.},
        title = "{Magnetic field amplification and magnetically supported explosions of collapsing, non-rotating stellar cores}",
      journal = {Mon. Not. Roy. Astron. Soc.},
         year = 2014,
        month = dec,
       volume = {445},
       number = {3},
        pages = {3169-3199},
          doi = {10.1093/mnras/stu1969},
archivePrefix = {arXiv},
       eprint = {1405.7466},
 primaryClass = {astro-ph.SR},
       adsurl = {https://ui.adsabs.harvard.edu/abs/2014MNRAS.445.3169O}
}

@ARTICLE{Wang2025,
       author = {{Wang}, Tianshu and {Burrows}, Adam},
        title = "{The Effect of the Fast-flavor Instability on Core-collapse Supernova Models}",
      journal = {Astrophys. J.},
         year = 2025,
        month = jun,
       volume = {986},
       number = {2},
          eid = {153},
        pages = {153},
          doi = {10.3847/1538-4357/add889},
archivePrefix = {arXiv},
       eprint = {2503.04896},
 primaryClass = {astro-ph.HE},
       url = {https://ui.adsabs.harvard.edu/abs/2025ApJ...986..153W}
}

@ARTICLE{Akaho2026,
    author = "Akaho, Ryuichiro and Nagakura, Hiroki and Iwakami, Wakana and Furusawa, Shun and Harada, Akira and Okawa, Hirotada and Matsufuru, Hideo and Sumiyoshi, Kohsuke and Yamada, Shoichi",
    title = "{Bifurcated Impact of Neutrino Fast Flavor Conversion on Core-Collapse Supernovae Informed by Multiangle Neutrino Radiation Hydrodynamics}",
    eprint = "2601.08269",
    archivePrefix = "arXiv",
    primaryClass = "astro-ph.HE",
    doi = "10.1103/fksy-1jtw",
    journal = "Phys. Rev. Lett.",
    volume = "136",
    number = "19",
    pages = "191002",
    year = "2026"
}

@article{Ehring2023abs,
    author = "Ehring, Jakob and Abbar, Sajad and Janka, Hans-Thomas and Raffelt, Georg G. and Tamborra, Irene",
    title = "{Fast Neutrino Flavor Conversions Can Help and Hinder Neutrino-Driven Explosions}",
    eprint = "2305.11207",
    archivePrefix = "arXiv",
    primaryClass = "astro-ph.HE",
    doi = "10.1103/PhysRevLett.131.061401",
    journal = "Phys. Rev. Lett.",
    volume = "131",
    number = "6",
    pages = "061401",
    year = "2023"
}

@ARTICLE{OConnor2010,
       author = {{O'Connor}, Evan and {Ott}, Christian D.},
        title = "{A new open-source code for spherically symmetric stellar collapse to neutron stars and black holes}",
      journal = {Class. Quant. Grav.},
         year = 2010,
        month = jun,
       volume = {27},
       number = {11},
          eid = {114103},
        pages = {114103},
          doi = {10.1088/0264-9381/27/11/114103},
archivePrefix = {arXiv},
       eprint = {0912.2393},
 primaryClass = {astro-ph.HE},
       url = {https://ui.adsabs.harvard.edu/abs/2010CQGra..27k4103O}
}

@ARTICLE{OConnor2015,
       author = {{O'Connor}, Evan},
        title = "{An Open-source Neutrino Radiation Hydrodynamics Code for Core-collapse Supernovae}",
      journal = {Astrophys. J. Supp.},
         year = 2015,
        month = aug,
       volume = {219},
       number = {2},
          eid = {24},
        pages = {24},
          doi = {10.1088/0067-0049/219/2/24},
archivePrefix = {arXiv},
       eprint = {1411.7058},
 primaryClass = {astro-ph.HE},
       url = {https://ui.adsabs.harvard.edu/abs/2015apjsup..219...24O}
}

@ARTICLE{Boccioli2021_STIR_GR,
        author = {{Boccioli}, Luca and {Mathews}, Grant J. and {O'Connor}, Evan P.},
        title = "{General Relativistic Neutrino-driven Turbulence in One-dimensional Core-collapse Supernovae}",
      journal = {Astrophys. J.},
         year = 2021,
        month = may,
       volume = {912},
       number = {1},
          eid = {29},
        pages = {29},
          doi = {10.3847/1538-4357/abe767},
archivePrefix = {arXiv},
       eprint = {2102.06767},
 primaryClass = {astro-ph.HE},
       url = {https://ui.adsabs.harvard.edu/abs/2021ApJ...912...29B}
}

@ARTICLE{Janka2025,
 author = "Janka, Hans-Thomas",
    title = "{Long-Term Multidimensional Models of Core-Collapse Supernovae: Progress and Challenges}",
    eprint = "2502.14836",
    archivePrefix = "arXiv",
    primaryClass = "astro-ph.HE",
    doi = "10.1146/annurev-nucl-121423-100945",
    journal = "Ann. Rev. Nucl. Part. Sci.",
    volume = "75",
    number = "1",
    pages = "425--461",
    year = "2025"
}

@ARTICLE{CARDALL2013,
       author = {{Cardall}, Christian Y. and {Endeve}, Eirik and {Mezzacappa}, Anthony},
        title = "{Conservative 3+1 general relativistic variable Eddington tensor radiation transport equations}",
      journal = {Phys. Rev.  D},
         year = 2013,
        month = may,
       volume = {87},
       number = {10},
          eid = {103004},
        pages = {103004},
          doi = {10.1103/PhysRevD.87.103004},
archivePrefix = {arXiv},
       eprint = {1209.2151},
 primaryClass = {astro-ph.HE},
       url = {https://ui.adsabs.harvard.edu/abs/2013PhRvD..87j3004C}
}

@ARTICLE{SHIBATA2011,
       author = {{Shibata}, M. and {Kiuchi}, K. and {Sekiguchi}, Y. and {Suwa}, Y.},
        title = "{Truncated Moment Formalism for Radiation Hydrodynamics in Numerical Relativity}",
      journal = {Prog. Theor. Phys.},
         year = 2011,
        month = jun,
       volume = {125},
       number = {6},
        pages = {1255-1287},
          doi = {10.1143/PTP.125.1255},
archivePrefix = {arXiv},
       eprint = {1104.3937},
 primaryClass = {astro-ph.HE},
       url = {https://ui.adsabs.harvard.edu/abs/2011PThPh.125.1255S}
}

@ARTICLE{Couch2020_STIR,
       author = {{Couch}, Sean M. and {Warren}, MacKenzie L. and {O'Connor}, Evan P.},
        title = "{Simulating Turbulence-aided Neutrino-driven Core-collapse Supernova Explosions in One Dimension}",
      journal = {Astrophys. J.},
         year = 2020,
        month = feb,
       volume = {890},
       number = {2},
          eid = {127},
        pages = {127},
          doi = {10.3847/1538-4357/ab609e},
archivePrefix = {arXiv},
       eprint = {1902.01340},
 primaryClass = {astro-ph.HE},
       url = {https://ui.adsabs.harvard.edu/abs/2020ApJ...890..127C}
}

@article{LATTIMER1991,
 author = "Lattimer, James M. and Swesty, F. Douglas",
    title = "{A Generalized equation of state for hot, dense matter}",
    doi = "10.1016/0375-9474(91)90452-C",
    journal = "Nucl. Phys. A",
    volume = "535",
    pages = "331--376",
    year = "1991"
}

@ARTICLE{Sukhbold2016,
       author = {{Sukhbold}, Tuguldur and {Ertl}, T. and {Woosley}, S.~E. and {Brown}, Justin M. and {Janka}, H.-T.},
        title = "{Core-collapse Supernovae from 9 to 120 Solar Masses Based on Neutrino-powered Explosions}",
      journal = {Astrophys. J.},
         year = 2016,
        month = apr,
       volume = {821},
       number = {1},
          eid = {38},
        pages = {38},
          doi = {10.3847/0004-637X/821/1/38},
archivePrefix = {arXiv},
       eprint = {1510.04643},
 primaryClass = {astro-ph.HE},
       url = {https://ui.adsabs.harvard.edu/abs/2016ApJ...821...38S}
}

@ARTICLE{Gogilashvili2022,
       author = {{Gogilashvili}, Mariam and {Murphy}, Jeremiah W.},
        title = "{A force explosion condition for spherically symmetric core-collapse supernovae}",
      journal = {Mon. Not.  Roy. Astron. Soc.},
         year = 2022,
        month = sep,
       volume = {515},
       number = {2},
        pages = {1610-1623},
          doi = {10.1093/mnras/stac1811},
archivePrefix = {arXiv},
       eprint = {2110.10173},
 primaryClass = {astro-ph.HE},
       url = {https://ui.adsabs.harvard.edu/abs/2022MNRAS.515.1610G}
}

@ARTICLE{Gogilashvili2023,
       author = {{Gogilashvili}, Mariam and {Murphy}, Jeremiah W. and {O'Connor}, Evan P.},
        title = "{The force explosion condition is consistent with spherically symmetric CCSN explosions}",
      journal = {Mon. Not.  Roy. Astron. Soc.},
         year = 2023,
        month = sep,
       volume = {524},
       number = {3},
        pages = {4109-4115},
          doi = {10.1093/mnras/stad2155},
archivePrefix = {arXiv},
       eprint = {2302.04890},
 primaryClass = {astro-ph.HE},
       url = {https://ui.adsabs.harvard.edu/abs/2023MNRAS.524.4109G}
}

@ARTICLE{Gogilashvili2024,
       author = {{Gogilashvili}, Mariam and {Murphy}, Jeremiah W. and {Miller}, Jonah M.},
        title = "{Including Neutrino-driven Convection in the Force Explosion Condition to Predict Explodability of Multidimensional Core-collapse Supernovae (FEC+)}",
      journal = {Astrophys. J.},
         year = 2024,
        month = feb,
       volume = {962},
       number = {2},
          eid = {110},
        pages = {110},
          doi = {10.3847/1538-4357/ad1d5e},
archivePrefix = {arXiv},
       eprint = {2311.02179},
 primaryClass = {astro-ph.HE},
       url = {https://ui.adsabs.harvard.edu/abs/2024ApJ...962..110G}
}

@ARTICLE{Boccioli2025,
       author = {{Boccioli}, Luca and {Gogilashvili}, Mariam and {Murphy}, Jeremiah and {O'Connor}, Evan P.},
        title = "{Quantifying the impact of the Si/O interface in CCSN explosions using the Force Explosion Condition}",
      journal = {Mon. Not.  Roy. Astron. Soc.},
         year = 2025,
        month = feb,
       volume = {537},
       number = {2},
        pages = {1182-1196},
          doi = {10.1093/mnras/staf066},
archivePrefix = {arXiv},
       eprint = {2410.17232},
 primaryClass = {astro-ph.HE},
       url = {https://ui.adsabs.harvard.edu/abs/2025MNRAS.537.1182B}
}

@ARTICLE{Burrows2021,
       author = {{Burrows}, A. and {Vartanyan}, D.},
        title = "{Core-collapse supernova explosion theory}",
      journal = {Nature},
         year = 2021,
        month = jan,
       volume = {589},
       number = {7840},
        pages = {29-39},
          doi = {10.1038/s41586-020-03059-w},
archivePrefix = {arXiv},
       eprint = {2009.14157},
 primaryClass = {astro-ph.SR},
       url = {https://ui.adsabs.harvard.edu/abs/2021Natur.589...29B}
}

@ARTICLE{Janka2016,
       author = {{Janka}, Hans-Thomas and {Melson}, Tobias and {Summa}, Alexander},
        title = "{Physics of Core-Collapse Supernovae in Three Dimensions: A Sneak Preview}",
      journal = {Ann. Rev. Nucl. Part. Sci.},
         year = 2016,
        month = oct,
       volume = {66},
       number = {1},
        pages = {341-375},
          doi = {10.1146/annurev-nucl-102115-044747},
archivePrefix = {arXiv},
       eprint = {1602.05576},
 primaryClass = {astro-ph.SR},
       url = {https://ui.adsabs.harvard.edu/abs/2016ARNPS..66..341J}
}

@ARTICLE{Mezzacappa2020,
       author = {{Mezzacappa}, Anthony and {Endeve}, Eirik and {Messer}, O.~E. Bronson and {Bruenn}, Stephen W.},
        title = "{Physical, numerical, and computational challenges of modeling neutrino transport in core-collapse supernovae}",
      journal = {Liv. Rev. Comput. Astrophys.},
         year = 2020,
        month = dec,
       volume = {6},
       number = {1},
          eid = {4},
        pages = {4},
          doi = {10.1007/s41115-020-00010-8},
archivePrefix = {arXiv},
       eprint = {2010.09013},
 primaryClass = {astro-ph.HE},
       url = {https://ui.adsabs.harvard.edu/abs/2020LRCA....6....4M}
}

@ARTICLE{Fischer2024,
       author = {{Fischer}, Tobias and {Guo}, Gang and {Langanke}, Karlheinz and {Mart{\'\i}nez-Pinedo}, Gabriel and {Qian}, Yong-Zhong and {Wu}, Meng-Ru},
        title = "{Neutrinos and nucleosynthesis of elements}",
      journal = {Prog. Part. Nucl. Phys.},
         year = 2024,
        month = may,
       volume = {137},
          eid = {104107},
        pages = {104107},
          doi = {10.1016/j.ppnp.2024.104107},
archivePrefix = {arXiv},
       eprint = {2308.03962},
 primaryClass = {astro-ph.HE},
       url = {https://ui.adsabs.harvard.edu/abs/2024PrPNP.13704107F}
}

@ARTICLE{Oertel2017,
       author = {{Oertel}, M. and {Hempel}, M. and {Kl{\"a}hn}, T. and {Typel}, S.},
        title = "{Equations of state for supernovae and compact stars}",
      journal = {Rev. Mod. Phys.},
         year = 2017,
        month = jan,
       volume = {89},
       number = {1},
          eid = {015007},
        pages = {015007},
          doi = {10.1103/RevModPhys.89.015007},
archivePrefix = {arXiv},
       eprint = {1610.03361},
 primaryClass = {astro-ph.HE},
       url = {https://ui.adsabs.harvard.edu/abs/2017RvMP...89a5007O}
}

@ARTICLE{Dasgupta2012,
       author = {{Dasgupta}, Basudeb and {O'Connor}, Evan P. and {Ott}, Christian D.},
        title = "{Role of collective neutrino flavor oscillations in core-collapse supernova shock revival}",
      journal = {Phys. Rev.  D},
         year = 2012,
        month = mar,
       volume = {85},
       number = {6},
          eid = {065008},
        pages = {065008},
          doi = {10.1103/PhysRevD.85.065008},
archivePrefix = {arXiv},
       eprint = {1106.1167},
 primaryClass = {astro-ph.SR},
       url = {https://ui.adsabs.harvard.edu/abs/2012PhRvD..85f5008D}
}

@ARTICLE{Strack2005,
       author = {{Strack}, P. and {Burrows}, A.},
        title = "{Generalized Boltzmann formalism for oscillating neutrinos}",
      journal = {Phys. Rev. D},
         year = 2005,
        month = may,
       volume = {71},
       number = {9},
          eid = {093004},
        pages = {093004},
          doi = {10.1103/PhysRevD.71.093004},
archivePrefix = {arXiv},
       eprint = {hep-ph/0504035},
 primaryClass = {hep-ph},
       url = {https://ui.adsabs.harvard.edu/abs/2005PhRvD..71i3004S}
}

@ARTICLE{Tamborra2021,
       author = {{Tamborra}, Irene and {Shalgar}, Shashank},
        title = "{New Developments in Flavor Evolution of a Dense Neutrino Gas}",
      journal = {Ann. Rev. Nucl. Part. Sci.},
         year = 2021,
        month = sep,
       volume = {71},
        pages = {165-188},
          doi = {10.1146/annurev-nucl-102920-050505},
archivePrefix = {arXiv},
       eprint = {2011.01948},
 primaryClass = {astro-ph.HE},
       url = {https://ui.adsabs.harvard.edu/abs/2021ARNPS..71..165T}
}

@ARTICLE{Johns2025,
       author = {{Johns}, Lucas and {Richers}, Sherwood and {Wu}, Meng-Ru},
        title = "{Neutrino Oscillations in Core-Collapse Supernovae and Neutron Star Mergers}",
      journal = {Ann. Rev. Nucl. Part. Sci.},
         year = 2025,
        month = sep,
       volume = {75},
       number = {1},
        pages = {399-423},
          doi = {10.1146/annurev-nucl-121423-100853},
archivePrefix = {arXiv},
       eprint = {2503.05959},
 primaryClass = {astro-ph.HE},
       url = {https://ui.adsabs.harvard.edu/abs/2025ARNPS..75..399J}
}

@ARTICLE{Volpe2024,
       author = {{Volpe}, M. Cristina},
        title = "{Neutrinos from dense environments: Flavor mechanisms, theoretical approaches, observations, and new directions}",
      journal = {Rev. Mod. Phys.},
         year = 2024,
        month = jun,
       volume = {96},
       number = {2},
          eid = {025004},
        pages = {025004},
          doi = {10.1103/RevModPhys.96.025004},
archivePrefix = {arXiv},
       eprint = {2301.11814},
 primaryClass = {hep-ph},
       url = {https://ui.adsabs.harvard.edu/abs/2024RvMP...96b5004V}
}

@inproceedings{Raffelt2025,
author = "Raffelt, Georg G. and Janka, Hans-Thomas and Fiorillo, Damiano F. G.",
    title = "{Neutrinos from core-collapse supernovae}",
    eprint = "2509.16306",
    archivePrefix = "arXiv",
    primaryClass = "astro-ph.HE",
    month = "9",
    year = "2025"
    }

@ARTICLE{Tamborra2025,
       author = {{Tamborra}, Irene},
        title = "{Neutrinos from explosive transients at the dawn of multi-messenger astronomy}",
      journal = {Nature Rev. Phys.},
         year = 2025,
        month = jun,
       volume = {7},
       number = {6},
        pages = {285-298},
          doi = {10.1038/s42254-025-00828-2},
archivePrefix = {arXiv},
       eprint = {2412.09699},
 primaryClass = {astro-ph.HE},
       url = {https://ui.adsabs.harvard.edu/abs/2025NatRP...7..285T}
}

@ARTICLE{Izaguirre2017,
       author = {{Izaguirre}, Ignacio and {Raffelt}, Georg G. and {Tamborra}, Irene},
        title = "{Fast Pairwise Conversion of Supernova Neutrinos: A Dispersion Relation Approach}",
      journal = {Phys. Rev. Lett.},
         year = 2017,
        month = jan,
       volume = {118},
       number = {2},
          eid = {021101},
        pages = {021101},
          doi = {10.1103/PhysRevLett.118.021101},
archivePrefix = {arXiv},
       eprint = {1610.01612},
 primaryClass = {hep-ph},
       url = {https://ui.adsabs.harvard.edu/abs/2017PhRvL.118b1101I}
}

@ARTICLE{Chakraborty2016,
       author = {{Chakraborty}, Sovan and {Hansen}, Rasmus and {Izaguirre}, Ignacio and {Raffelt}, Georg G.},
        title = "{Collective neutrino flavor conversion: Recent developments}",
      journal = {Nuc. Phys. B},
         year = 2016,
        month = jul,
       volume = {908},
        pages = {366-381},
          doi = {10.1016/j.nuclphysb.2016.02.012},
archivePrefix = {arXiv},
       eprint = {1602.02766},
 primaryClass = {hep-ph},
       url = {https://ui.adsabs.harvard.edu/abs/2016NuPhB.908..366C}
}

@ARTICLE{Shalgar2023,
       author = {{Shalgar}, Shashank and {Tamborra}, Irene},
        title = "{Neutrino decoupling is altered by flavor conversion}",
      journal = {Phys. Rev.  D},
         year = 2023,
        month = aug,
       volume = {108},
       number = {4},
          eid = {043006},
        pages = {043006},
          doi = {10.1103/PhysRevD.108.043006},
archivePrefix = {arXiv},
       eprint = {2206.00676},
 primaryClass = {astro-ph.HE},
       url = {https://ui.adsabs.harvard.edu/abs/2023PhRvD.108d3006S}
}

@ARTICLE{Shalgar2023a,
       author = {{Shalgar}, Shashank and {Tamborra}, Irene},
        title = "{Neutrino flavor conversion, advection, and collisions: Toward the full solution}",
      journal = {Phys. Rev.  D},
         year = 2023,
        month = mar,
       volume = {107},
       number = {6},
          eid = {063025},
        pages = {063025},
          doi = {10.1103/PhysRevD.107.063025},
archivePrefix = {arXiv},
       eprint = {2207.04058},
 primaryClass = {astro-ph.HE},
       url = {https://ui.adsabs.harvard.edu/abs/2023PhRvD.107f3025S}
}

@ARTICLE{Johns2023,
       author = {{Johns}, Lucas},
        title = "{Collisional Flavor Instabilities of Supernova Neutrinos}",
      journal = {Phys. Rev. Lett.},
         year = 2023,
        month = may,
       volume = {130},
       number = {19},
          eid = {191001},
        pages = {191001},
          doi = {10.1103/PhysRevLett.130.191001},
archivePrefix = {arXiv},
       eprint = {2104.11369},
 primaryClass = {hep-ph},
       url = {https://ui.adsabs.harvard.edu/abs/2023PhRvL.130s1001J}
}

@ARTICLE{Nagakura2022,
       author = {{Nagakura}, Hiroki},
        title = "{General-relativistic quantum-kinetics neutrino transport}",
      journal = {Phys. Rev.  D},
         year = 2022,
        month = sep,
       volume = {106},
       number = {6},
          eid = {063011},
        pages = {063011},
          doi = {10.1103/PhysRevD.106.063011},
archivePrefix = {arXiv},
       eprint = {2206.04098},
 primaryClass = {astro-ph.HE},
       url = {https://ui.adsabs.harvard.edu/abs/2022PhRvD.106f3011N}
}

@ARTICLE{Fiorillo2025,
       author = {{Fiorillo}, Damiano F.~G. and {Janka}, Hans-Thomas and {Raffelt}, Georg G.},
        title = "{Neutrino-Mass-Driven Instabilities as the Earliest Flavor Conversion in Supernovae}",
      journal = {Phys. Rev. Lett.},
         year = 2025,
        month = dec,
       volume = {135},
       number = {23},
          eid = {231003},
        pages = {231003},
          doi = {10.1103/jbmx-rbzt},
archivePrefix = {arXiv},
       eprint = {2507.22985},
 primaryClass = {hep-ph},
       url = {https://ui.adsabs.harvard.edu/abs/2025PhRvL.135w1003F}
}

@ARTICLE{Xiong2024,
       author = {{Xiong}, Zewei and {Wu}, Meng-Ru and {George}, Manu and {Lin}, Chun-Yu and {Khosravi Largani}, Noshad and {Fischer}, Tobias and {Mart{\'\i}nez-Pinedo}, Gabriel},
        title = "{Fast neutrino flavor conversions in a supernova: Emergence, evolution, and effects}",
      journal = {Phys. Rev.  D},
         year = 2024,
        month = jun,
       volume = {109},
       number = {12},
          eid = {123008},
        pages = {123008},
          doi = {10.1103/PhysRevD.109.123008},
archivePrefix = {arXiv},
       eprint = {2402.19252},
 primaryClass = {astro-ph.HE},
       url = {https://ui.adsabs.harvard.edu/abs/2024PhRvD.109l3008X}
}

@ARTICLE{Cornelius2024,
       author = {{Cornelius}, Marie and {Shalgar}, Shashank and {Tamborra}, Irene},
        title = "{Neutrino quantum kinetics in two spatial dimensions}",
      journal = {JCAP},
         year = 2024,
        month = nov,
       volume = {2024},
       number = {11},
          eid = {060},
        pages = {060},
          doi = {10.1088/1475-7516/2024/11/060},
archivePrefix = {arXiv},
       eprint = {2407.04769},
 primaryClass = {astro-ph.HE},
       url = {https://ui.adsabs.harvard.edu/abs/2024JCAP...11..060C}
}

@ARTICLE{Shalgar2025,
       author = {{Shalgar}, Shashank and {Tamborra}, Irene},
        title = "{Neutrino quantum kinetics in three flavors}",
      journal = {JCAP},
         year = 2025,
        month = dec,
       volume = {2025},
       number = {12},
          eid = {026},
        pages = {026},
          doi = {10.1088/1475-7516/2025/12/026},
archivePrefix = {arXiv},
       eprint = {2503.03835},
 primaryClass = {astro-ph.HE},
       url = {https://ui.adsabs.harvard.edu/abs/2025JCAP...12..026S}
}

@ARTICLE{Just2022,
       author = {{Just}, Oliver and {Abbar}, Sajad and {Wu}, Meng-Ru and {Tamborra}, Irene and {Janka}, Hans-Thomas and {Capozzi}, Francesco},
        title = "{Fast neutrino conversion in hydrodynamic simulations of neutrino-cooled accretion disks}",
      journal = {Phys. Rev.  D},
         year = 2022,
        month = apr,
       volume = {105},
       number = {8},
          eid = {083024},
        pages = {083024},
          doi = {10.1103/PhysRevD.105.083024},
archivePrefix = {arXiv},
       eprint = {2203.16559},
 primaryClass = {astro-ph.HE},
       url = {https://ui.adsabs.harvard.edu/abs/2022PhRvD.105h3024J}
}

@ARTICLE{Goimil2025,
       author = {{Goimil-Garc{\'\i}a}, Manuel and {Tamborra}, Irene},
        title = "{Steady state of fast-oscillating neutrinos in an inhomogeneous medium}",
      journal = {Phys. Rev. D},
         year = 2025,
        month = nov,
       volume = {112},
       number = {10},
          eid = {103011},
        pages = {103011},
          doi = {10.1103/gdg9-rzns},
archivePrefix = {arXiv},
       eprint = {2509.22805},
 primaryClass = {astro-ph.HE},
       url = {https://ui.adsabs.harvard.edu/abs/2025PhRvD.112j3011G}
}

@ARTICLE{Padilla2022,
       author = {{Padilla-Gay}, Ian and {Tamborra}, Irene and {Raffelt}, Georg G.},
        title = "{Neutrino Flavor Pendulum Reloaded: The Case of Fast Pairwise Conversion}",
      journal = {Phys. Rev. Lett.},
         year = 2022,
        month = mar,
       volume = {128},
       number = {12},
          eid = {121102},
        pages = {121102},
          doi = {10.1103/PhysRevLett.128.121102},
archivePrefix = {arXiv},
       eprint = {2109.14627},
 primaryClass = {astro-ph.HE},
       url = {https://ui.adsabs.harvard.edu/abs/2022PhRvL.128l1102P}
}

@ARTICLE{Liu2025,
       author = {{Liu}, Jiabao and {Nagakura}, Hiroki and {Zaizen}, Masamichi and {Johns}, Lucas and {Yamada}, Shoichi},
        title = "{Asymptotic states of fast neutrino-flavor conversions in the three-flavor framework}",
      journal = {Phys. Rev.  D},
         year = 2025,
        month = jun,
       volume = {111},
       number = {12},
          eid = {123004},
        pages = {123004},
          doi = {10.1103/v9lr-ydbb},
archivePrefix = {arXiv},
       eprint = {2503.18145},
 primaryClass = {astro-ph.HE},
       url = {https://ui.adsabs.harvard.edu/abs/2025PhRvD.111l3004L}
}

@ARTICLE{Zaizen2023,
       author = {{Zaizen}, Masamichi and {Nagakura}, Hiroki},
        title = "{Simple method for determining asymptotic states of fast neutrino-flavor conversion}",
      journal = {Phys. Rev.  D},
         year = 2023,
        month = may,
       volume = {107},
       number = {10},
          eid = {103022},
        pages = {103022},
          doi = {10.1103/PhysRevD.107.103022},
archivePrefix = {arXiv},
       eprint = {2211.09343},
 primaryClass = {astro-ph.HE},
       url = {https://ui.adsabs.harvard.edu/abs/2023PhRvD.107j3022Z}
}

@ARTICLE{Zaizen2023a,
       author = {{Zaizen}, Masamichi and {Nagakura}, Hiroki},
        title = "{Characterizing quasisteady states of fast neutrino-flavor conversion by stability and conservation laws}",
      journal = {Phys. Rev.  D},
         year = 2023,
        month = jun,
       volume = {107},
       number = {12},
          eid = {123021},
        pages = {123021},
          doi = {10.1103/PhysRevD.107.123021},
archivePrefix = {arXiv},
       eprint = {2304.05044},
 primaryClass = {astro-ph.HE},
       url = {https://ui.adsabs.harvard.edu/abs/2023PhRvD.107l3021Z}
}

@ARTICLE{Xiong2021,
       author = {{Xiong}, Zewei and {Qian}, Yong-Zhong},
        title = "{Stationary solutions for fast flavor oscillations of a homogeneous dense neutrino gas}",
      journal = {Phys. Lett. B},
         year = 2021,
        month = sep,
       volume = {820},
          eid = {136550},
        pages = {136550},
          doi = {10.1016/j.physletb.2021.136550},
archivePrefix = {arXiv},
       eprint = {2104.05618},
 primaryClass = {astro-ph.HE},
       url = {https://ui.adsabs.harvard.edu/abs/2021PhLB..82036550X}
}

@ARTICLE{Xiong2025,
       author = {{Xiong}, Zewei and {Wu}, Meng-Ru and {George}, Manu and {Lin}, Chun-Yu},
        title = "{Robust Integration of Fast Flavor Conversions in Classical Neutrino Transport}",
      journal = {Phys. Rev. Lett.},
         year = 2025,
        month = feb,
       volume = {134},
       number = {5},
          eid = {051003},
        pages = {051003},
          doi = {10.1103/PhysRevLett.134.051003},
archivePrefix = {arXiv},
       eprint = {2403.17269},
 primaryClass = {astro-ph.HE},
       url = {https://ui.adsabs.harvard.edu/abs/2025PhRvL.134e1003X}
}

@ARTICLE{Johns2025_subgrid,
       author = {{Johns}, Lucas},
        title = "{Subgrid modeling of neutrino oscillations in astrophysics}",
      journal = {Phys. Rev.  D},
         year = 2025,
        month = aug,
       volume = {112},
       number = {4},
          eid = {043024},
        pages = {043024},
          doi = {10.1103/3fr2-qttd},
archivePrefix = {arXiv},
       eprint = {2401.15247},
 primaryClass = {astro-ph.HE},
       url = {https://ui.adsabs.harvard.edu/abs/2025PhRvD.112d3024J}
}

@ARTICLE{Nagakura2024,
       author = {{Nagakura}, Hiroki and {Johns}, Lucas and {Zaizen}, Masamichi},
        title = "{Bhatnagar-Gross-Krook subgrid model for neutrino quantum kinetics}",
      journal = {Phys. Rev.  D},
         year = 2024,
        month = apr,
       volume = {109},
       number = {8},
          eid = {083013},
        pages = {083013},
          doi = {10.1103/PhysRevD.109.083013},
archivePrefix = {arXiv},
       eprint = {2312.16285},
 primaryClass = {astro-ph.HE},
       url = {https://ui.adsabs.harvard.edu/abs/2024PhRvD.109h3013N}
}

@ARTICLE{Nagakura2023,
       author = {{Nagakura}, Hiroki},
        title = "{Roles of Fast Neutrino-Flavor Conversion on the Neutrino-Heating Mechanism of Core-Collapse Supernova}",
      journal = {Phys. Rev. Lett.},
         year = 2023,
        month = may,
       volume = {130},
       number = {21},
          eid = {211401},
        pages = {211401},
          doi = {10.1103/PhysRevLett.130.211401},
archivePrefix = {arXiv},
       eprint = {2301.10785},
 primaryClass = {astro-ph.HE},
       url = {https://ui.adsabs.harvard.edu/abs/2023PhRvL.130u1401N}
}

@ARTICLE{Padilla2025,
       author = {{Padilla-Gay}, Ian and {Chen}, Heng-Hao and {Abbar}, Sajad and {Wu}, Meng-Ru and {Xiong}, Zewei},
        title = "{Flavor equilibration of supernova neutrinos: Exploring the dynamics of slow modes}",
      journal = {Phys. Rev.  D},
         year = 2025,
        month = aug,
       volume = {112},
       number = {4},
          eid = {043039},
        pages = {043039},
          doi = {10.1103/jg14-8p4l},
archivePrefix = {arXiv},
       eprint = {2505.11588},
 primaryClass = {astro-ph.HE},
       url = {https://ui.adsabs.harvard.edu/abs/2025PhRvD.112d3039P}
}

@ARTICLE{Cornelius2025a,
       author = {{Cornelius}, Marie and {Tamborra}, Irene and {Heinlein}, Malte and {Janka}, Hans-Thomas},
        title = "{Diagnosing electron-neutrino lepton number crossings in core-collapse supernovae: A comparison of methods}",
      journal = {Phys. Rev.  D},
         year = 2025,
        month = sep,
       volume = {112},
       number = {6},
          eid = {063004},
        pages = {063004},
          doi = {10.1103/gqd7-4ynz},
archivePrefix = {arXiv},
       eprint = {2506.20723},
 primaryClass = {astro-ph.HE},
       url = {https://ui.adsabs.harvard.edu/abs/2025PhRvD.112f3004C}
}

@ARTICLE{Tamborra2017,
       author = {{Tamborra}, Irene and {H{\"u}depohl}, Lorenz and {Raffelt}, Georg G. and {Janka}, Hans-Thomas},
        title = "{Flavor-dependent Neutrino Angular Distribution in Core-collapse Supernovae}",
      journal = {Astrophys. J.},
         year = 2017,
        month = apr,
       volume = {839},
       number = {2},
          eid = {132},
        pages = {132},
          doi = {10.3847/1538-4357/aa6a18},
archivePrefix = {arXiv},
       eprint = {1702.00060},
 primaryClass = {astro-ph.HE},
       url = {https://ui.adsabs.harvard.edu/abs/2017ApJ...839..132T}
}

@ARTICLE{Sumiyoshi2012,
       author = {{Sumiyoshi}, K. and {Yamada}, S.},
        title = "{Neutrino Transfer in Three Dimensions for Core-collapse Supernovae. I. Static Configurations}",
      journal = {Astrophys. J. Supp.},
         year = 2012,
        month = mar,
       volume = {199},
       number = {1},
          eid = {17},
        pages = {17},
          doi = {10.1088/0067-0049/199/1/17},
archivePrefix = {arXiv},
       eprint = {1201.2244},
 primaryClass = {astro-ph.HE},
       url = {https://ui.adsabs.harvard.edu/abs/2012ApJS..199...17S}
}

@ARTICLE{Akaho2021,
       author = {{Akaho}, Ryuichiro and {Harada}, Akira and {Nagakura}, Hiroki and {Sumiyoshi}, Kohsuke and {Iwakami}, Wakana and {Okawa}, Hirotada and {Furusawa}, Shun and {Matsufuru}, Hideo and {Yamada}, Shoichi},
        title = "{Multidimensional Boltzmann Neutrino Transport Code in Full General Relativity for Core-collapse Simulations}",
      journal = {Astrophys. J.},
         year = 2021,
        month = mar,
       volume = {909},
       number = {2},
          eid = {210},
        pages = {210},
          doi = {10.3847/1538-4357/abe1bf},
archivePrefix = {arXiv},
       eprint = {2010.10780},
 primaryClass = {astro-ph.HE},
       url = {https://ui.adsabs.harvard.edu/abs/2021ApJ...909..210A}
}

@ARTICLE{Nagakura2018,
       author = {{Nagakura}, Hiroki and {Iwakami}, Wakana and {Furusawa}, Shun and {Okawa}, Hirotada and {Harada}, Akira and {Sumiyoshi}, Kohsuke and {Yamada}, Shoichi and {Matsufuru}, Hideo and {Imakura}, Akira},
        title = "{Simulations of Core-collapse Supernovae in Spatial Axisymmetry with Full Boltzmann Neutrino Transport}",
      journal = {Astrophys. J.},
         year = 2018,
        month = feb,
       volume = {854},
       number = {2},
          eid = {136},
        pages = {136},
          doi = {10.3847/1538-4357/aaac29},
archivePrefix = {arXiv},
       eprint = {1702.01752},
 primaryClass = {astro-ph.HE},
       url = {https://ui.adsabs.harvard.edu/abs/2018ApJ...854..136N}
}

@ARTICLE{Fischer2010,
       author = {{Fischer}, T. and {Whitehouse}, S.~C. and {Mezzacappa}, A. and {Thielemann}, F.-K. and {Liebend{\"o}rfer}, M.},
        title = "{Protoneutron star evolution and the neutrino-driven wind in general relativistic neutrino radiation hydrodynamics simulations}",
      journal = {Astron. Astrophys.},
         year = 2010,
        month = jul,
       volume = {517},
          eid = {A80},
        pages = {A80},
          doi = {10.1051/0004-6361/200913106},
archivePrefix = {arXiv},
       eprint = {0908.1871},
 primaryClass = {astro-ph.HE},
       url = {https://ui.adsabs.harvard.edu/abs/2010A&A...517A..80F}
}

@ARTICLE{Johns2021,
       author = {{Johns}, Lucas and {Nagakura}, Hiroki},
        title = "{Fast flavor instabilities and the search for neutrino angular crossings}",
      journal = {Phys. Rev.  D},
         year = 2021,
        month = jun,
       volume = {103},
       number = {12},
          eid = {123012},
        pages = {123012},
          doi = {10.1103/PhysRevD.103.123012},
archivePrefix = {arXiv},
       eprint = {2104.04106},
 primaryClass = {hep-ph},
       url = {https://ui.adsabs.harvard.edu/abs/2021PhRvD.103l3012J}
}

@ARTICLE{Shalgar2024,
       author = {{Shalgar}, Shashank and {Tamborra}, Irene},
        title = "{Neutrino quantum kinetics in a core-collapse supernova}",
      journal = {JCAP},
         year = 2024,
        month = sep,
       volume = {2024},
       number = {9},
          eid = {021},
        pages = {021},
          doi = {10.1088/1475-7516/2024/09/021},
archivePrefix = {arXiv},
       eprint = {2406.09504},
 primaryClass = {astro-ph.HE},
       url = {https://ui.adsabs.harvard.edu/abs/2024JCAP...09..021S}
}

@ARTICLE{Mori2025,
       author = {{Mori}, Kanji and {Takiwaki}, Tomoya and {Kotake}, Kei and {Horiuchi}, Shunsaku},
        title = "{Three-dimensional core-collapse supernova models with phenomenological treatment of neutrino flavor conversions}",
      journal = {Publ. Astron. Soc. Jap.},
         year = 2025,
        month = apr,
       volume = {77},
       number = {2},
        pages = {L9-L15},
          doi = {10.1093/pasj/psaf007},
archivePrefix = {arXiv},
       eprint = {2501.15256},
 primaryClass = {astro-ph.HE},
       url = {https://ui.adsabs.harvard.edu/abs/2025PASJ...77L...9M}
}

@ARTICLE{Boccioli2023_explodability,
       author = {{Boccioli}, Luca and {Roberti}, Lorenzo and {Limongi}, Marco and {Mathews}, Grant J. and {Chieffi}, Alessandro},
        title = "{Explosion Mechanism of Core-collapse Supernovae: Role of the Si/Si-O Interface}",
      journal = {Astrophys. J.},
         year = 2023,
        month = may,
       volume = {949},
       number = {1},
          eid = {17},
        pages = {17},
          doi = {10.3847/1538-4357/acc06a},
archivePrefix = {arXiv},
       eprint = {2207.08361},
 primaryClass = {astro-ph.HE},
       url = {https://ui.adsabs.harvard.edu/abs/2023ApJ...949...17B}
}

@ARTICLE{Wang2022_prog_study_ram_pressure,
       author = {{Wang}, Tianshu and {Vartanyan}, David and {Burrows}, Adam and {Coleman}, Matthew S.~B.},
        title = "{The essential character of the neutrino mechanism of core-collapse supernova explosions}",
      journal = {Mon. Not.  Roy. Astron. Soc.},
         year = 2022,
        month = nov,
       volume = {517},
       number = {1},
        pages = {543-559},
          doi = {10.1093/mnras/stac2691},
archivePrefix = {arXiv},
       eprint = {2207.02231},
 primaryClass = {astro-ph.SR},
       url = {https://ui.adsabs.harvard.edu/abs/2022MNRAS.517..543W}
}

@article{Gogilashvili2026_prl,
    author = "Gogilashvili, Mariam and Tamborra, Irene",
    title = "{Neutrino Flavor Conversion Shapes the Rate of Failed Core-collapse Supernovae}",
    journal = "Phys. Rev. D Letters, in press",
    year = "2026",
archivePrefix = {arXiv},
    eprint = {2605.16504},
primaryClass = {astro-ph.HE},
}

\end{document}